\documentclass[onecolumn,nofootinbib,showkeys]{revtex4}
\usepackage{setspace}
\usepackage{amsmath,amssymb}
\usepackage{graphicx}
\usepackage{dcolumn}
\usepackage{bm,color}
\usepackage{hyperref}
\usepackage{accents}
\usepackage{amssymb,float}
\usepackage{amsmath}
\usepackage{multirow}
\usepackage{tikz}
\usepackage{enumerate}
\usepackage{tensor}

\hypersetup{
    colorlinks=true,
    linkcolor=red,
    filecolor=magenta,      
    citecolor=blue
}

\begin{document}

\title{Gravitational Collapse of Anisotropic Compact Stars in Modified $ f(R) $ Gravity}

\author{Jay Solanki}
\email{jay565109@gmail.com}
\affiliation{Sardar Vallabhbhai National Institute of Technology,\\ Surat - 395007, Gujarat, India}

\begin{abstract}
    The physically realistic model of compact stars undergoing gravitational collapse in $ f(R) $ gravity has been developed. We consider a more general model $ R + f(R) = R + k R^m $ and describe the interior space-time of gravitationally collapsing stars with separable-form of metric admitting homothetic killing vector. We then investigate the junction conditions to match the interior space-time with exterior space-time. Considering all junction conditions, we find analytical solutions describing interior space-time metric, energy density, pressures, and heat flux density of the compact stars undergoing gravitational collapse. We impose the energy conditions to the model for describing the realistic collapse of physically possible matter distribution for particular models of $GR$, $ R+k R^2 $ and $ R+k R^4 $ gravity. The comprehensive graphical analysis of all energy conditions show that the model is physically acceptable and realistic. We additionally investigate the physical properties of collapsing stars which are useful to decipher the inherent nature of such gravitationally collapsing stars. 
\end{abstract}


\maketitle

\section{Introduction}
\label{sec1}
Einstein's theory of general relativity has been extensively used to describe various astronomical phenomena such as perihelion of mercury, various properties of compact stars and their gravitational collapse, formation of black holes, Etc. General relativity is found to be very successful in describing such phenomena with great accuracy. However, it fails to describe the observations of galaxy rotation curves and expansion of the universe-like phenomena. Such phenomena can be described with general relativity by introducing dark matter and dark energy\cite{ARUN2017166, sami2009dark, sami2009primer, COPELAND_2006, 2011AdAst2011E...8G}. Another possible solution is to modify the theory of general relativity.\cite{Nojiri:2003ft, Nojiri:2010wj, Nojiri:2017ncd,doi:10.1142/S0219887814600068,doi:10.1142/S0219887815300032,doi:10.1142/S0219887820500164,Capozziello_2011,durrer2008dark,RevModPhys.82.451} One possible modification in general relativity can be achieved by modifying the Einstein-Hilbert action with additional non-linear form $ f(R) $ of the Ricci scalar $ R $. Thus, the Lagrangian density will become $ R + f(R) $, apart from the Lagrangian density for the matter distribution. Here in this paper, we consider the general non-linear form of $ f(R) = k R^m $ giving modified gravity model $ R + f(R) = R + k R^m $. 
\par Modified $ f(R) $ gravity has been extensively used to describe various astronomical phenomena such as stable neutron stars and strange stars, the gravitational collapse of such compact stars and wormholes, galaxy rotation curves, and expansion of the universe. Many authors have investigated useful forms of $ f(R) $ and their validity to pass solar tests and cosmological bounds in describing various astronomical phenomena\cite{PhysRevD.78.064019, 2012,PhysRevD.83.064004,PhysRevD.85.044022}. There have been also huge efforts made by various authors to investigate compact stars and their gravitational collapse in $ f(R) $ gravity\cite{2011b,2010,Abbas:2017kra,article,PhysRevD.85.063518}. The authors of the paper \cite{PhysRevD.93.023501} have investigated the mass-radius relationship of neutron stars in $ f(R) $ gravity by specifying the equation of state relating the isotropic matter pressure and density. Extreme neutron stars have been investigated in the extended theory of gravity in the reference \cite{Astashenok_2013,Astashenok_2015}. Recently authors have studied super-massive neutron stars, and the causal limit of neutron star maximum mass in View of $GW190814$ \cite{2020c, 2021}. The authors of the paper \cite{2021c} have investigated novel stellar astrophysics from extended gravity. Neutron stars in scalar-tensor gravity with Higgs scalar potential are studied in the paper \cite{2021arXiv210401982O}. The gravitational collapse in modified $ f(R) $ gravity has also been studied in the context of cosmology by the authors of the paper\cite{PhysRevD.88.084015}. The spherical collapse of compact stars in $f(R)$ gravity and the Belinskii-Khalatnikov-Lifshitz conjecture has been investigated in the paper\cite{PhysRevD.90.024017}. Spherically symmetric collapse in modified $ f(R) $ gravity and a generalization to conformally flat stars are studied in \cite{2014,2018}. Gravitational collapse in generalized teleparallel gravity has been investigated in \cite{articlex}. Dynamics of the Charged Radiating and dissipative Collapses in Modified Gauss-Bonnet Gravity have been studied in these papers \cite{articlew,Abbas:2018ica}. Authors of the paper \cite{2021y} have studied dynamical conditions and causal transport of dissipative spherical collapse in f(R,T) gravity.
\par Many authors have also investigated the phenomena of gravitational collapse in $ f(R) $ gravity and the stability of polytropic fluid in $ f(R) $ gravity\cite{2019,2020a}. Gravitational collapse in repulsive $R+\mu^{4}/R$ gravity has been studied in the paper \cite{2016x}. The authors of the paper \cite{article6} have investigated the gravitational collapse of compact stars for charged adiabatic LTB configuration. Higher-dimensional charged LTB collapse in $f(R)$ gravity has been studied in reference \cite{articlez}. Recently, authors have investigated three different $ f(R) $ gravity models to study the collapsing phenomena as well as the nature of central singularity\cite{Jaryal:2021lsu}.  In the paper \cite{2008GReGr..40.2149P}, authors have studied radiating gravitational collapse considering the effect of shear viscosity. The gravitational collapse in $ f(R) $ gravity and method of $ R $ matching has been investigated in the paper \cite{2020b}. In that paper, it was found that the separable form of the Lemaitre-Tolman-Bondi metric is not very suitable for describing gravitational collapse in $ f(R) $ gravity due to violations of some junction conditions. However, recently it was found that a particular metric form admitting homothetic killing vector and non-separable of LTB metric satisfies all necessary junction conditions to describe realistic gravitational collapse. However, such detailed studies have been performed only for some restricted class of $ f(R) $ models like modified $ R^2 $ gravity. It is useful to investigate such physically important scenarios for a more general $ f(R) $ model to understand the inherent nature of gravitationally collapsing stars described in modified $ f(R) $ gravity. Thus, this paper studies the gravitationally collapsing stars in a more general $ R + f(R) = R + k R^m $ model.      
\par To investigate the gravitational collapse of compact stars in modified $ f(R) $ gravity, firstly, it is necessary to formulate field equations of $ f(R) $ gravity. Varying the modified Einstein-Hilbert action in terms of modified Lagrangian density  $ R + f(R) + \mathcal{L}_{matter} $, the field equations of modified $ f(R) $ gravity is being obtained. Then the field equations have to be solved for particular forms of interior space-time metric representing the gravitational collapse of compact stars. While solving field equations of modified $ f(R) $ gravity, junction conditions must be considered for smooth matching of interior and exterior space-time. In general relativity, these boundary conditions are known as the Darmois-Israel conditions. These conditions require the first and second fundamental forms to be matched on the boundary of collapsing stars. However, in $ f(R) $ gravity, extra junction conditions have to be applied to smoothly match the interior and exterior space-time\cite{10.1143/PTP.119.237,PhysRevD.88.064015}.   These extra junction conditions admitting strong boundary conditions greatly restrict the physically possible solutions of the problem. After solving field equations of gravitational collapse in $ f(R) $ gravity, the solutions have to be satisfied the energy conditions for the description of physically possible matter distribution undergoing gravitational collapse. Finally, important physical properties of compact stars undergoing gravitational collapse can be investigated from the generated physically well-behaved model.     
\par This paper is organized as follows. In section \ref{sec2}, we describe the field equations of gravitationally collapsing stars in $ f(R) $ gravity for the specific interior metric admitting homothetic killing vector. Then in section \ref{sec3}, we develop junction conditions for the form of interior space-time under consideration and exterior space-time described by the Vaidya metric. We investigate the conditions for smooth matching of interior and exterior metrics, their exterior derivatives, and Ricci scalars and their derivatives. In section \ref{sec4}, we solve field equations by using the method of $ R $ matching to find exact solutions describing gravitationally collapsing stars and their physical quantities for more general model $ f(R) = k R^m $. Then in section \ref{sec5}, we describe necessary energy conditions to be imposed for obtaining a physically acceptable model of gravitational collapse in $ f(R) $ gravity. In \ref{sec6}, we do physical analysis of solutions for particular models of $ GR $ by choosing parameter $ k = 0 $, $ R + f(R) = R + k R^2 $ by choosing parameter $ m = 2 $ and  $ R + f(R) = R + k R^4 $ gravity by choosing parameter $ m = 4 $. We find model parameters such that the models satisfy all energy conditions, and we also plot all energy conditions for each model. The graphs of all energy conditions plotted for each model show that models are physically acceptable. In section \ref{sec7} we investigate the physical properties of gravitationally collapsing stars and summarize all the results in the Discussion section \ref{sec8}.

\section{Field equations of gravitational collapse in $ f(R) $ gravity}
\label{sec2}
The field equations of $ f(R) $ gravity can be obtained by modifying the Einstein-Hilbert action. The Einstein-Hilbert action can be modified by introducing modification in terms of the additional function of curvature $ f(R) $. Thus, the Lagrangian density for modified $ f(R) $ gravity becomes $ R + f(R) + \mathcal{L}_{matter} $. Variation of the action written in terms of modified Lagrangian density yields the following field equation of modified $ f(R) $ gravity

\begin{equation}
    \label{1}
    G_{\alpha \beta} = \frac{1}{1+f_R}\left( T_{\alpha \beta} + \nabla_{\alpha}\nabla_{\beta}f_R - g_{\alpha \beta}\nabla_{\mu}\nabla^{\mu}f_R + \frac{1}{2}g_{\alpha \beta}(f - R f_R) \right)
\end{equation}

Where $ G_{\alpha \beta} = R_{\alpha \beta} - \frac{1}{2}g_{\alpha \beta}R$ is the well known Einstein tensor. $ T_{\alpha \beta} $ is the energy-momentum tensor and $ f_R = \frac{df(R)}{dR} $. The equation (\ref{1}) can be written in the following form 

\begin{equation}
    \label{2}
    G_{\alpha \beta} = \frac{1}{1+f_R}\left( T^m_{\alpha \beta} + T^c_{\alpha \beta} \right)
\end{equation}

Where the $ T^m_{\alpha \beta} $ is the energy-momentum tensor for the matter and the $ T^c_{\alpha \beta}$ is known as effective energy-momentum tensor written in terms of the function of Ricci scalar and its derivatives. Effective energy-momentum tensor is the additional source of the curvature in the modified $ f(R) $ gravity, and thus it is of purely geometrical origin. To study the spherical gravitationally collapsing stars, we consider the following form of energy-momentum tensor for anisotropic matter distribution

\begin{equation}
    \label{3}
    T_{\alpha \beta}^m = (\rho + p_t)u_{\alpha}u_{\beta} + p_t g_{\alpha \beta} - (p_t - p_r)v_{\alpha}v_{\beta} + 2qu_{(\alpha} v_{{\beta})}
\end{equation}
 
Where $ \rho $, $ p_r $, $ p_t $ and $ q $ denotes the energy density of matter, radial and tangential pressure and heat flux density respectively. $ u^{\alpha} $ and $ v^{\alpha} $ denotes the four velocity and vector in radial direction respectively. The vector representing heat flux can be written as $ q^{\alpha} = q v^{\alpha} $. All these three vectors follows the normalisation conditions given by $ u^{\alpha}u_{\alpha} = -1 $ ,$ v^{\alpha}v_{\alpha} = 1 $ and $ u^{\alpha}q_{\alpha} = 0 $.
\par From equation (\ref{1}), the effective energy-momentum tensor $ T^C_{\alpha \beta} $ is given by

\begin{equation}
    \label{4}
    T^c_{\alpha \beta} = \nabla_{\alpha}\nabla_{\beta}f_R - g_{\alpha \beta}\nabla_{\mu}\nabla^{\mu}f_R + \frac{1}{2}g_{\alpha \beta}(f - R f_R)
\end{equation}

To study the gravitational collapse of spherically symmetric compact stars in $ f(R) $ gravity, we use the following form of space-time metric
\begin{equation}
    \label{5}
    ds^2 = - y(r)^2dt^2 + 2c(t)^2\left(\frac{dy}{dr}\right)^2dr^2 + c(t)^2y(r)^2(d\theta^2 + (sin^2\theta) d\phi^2)
\end{equation}

From equation (\ref{1}), we obtain independent field equations of gravitational collapse in $ f(R) $ gravity for metric (\ref{1}) as follow, (in units of $ c = 8\pi G = 1 $)

\begin{equation}
    \label{6}
     \rho y(r)^2 = \left(\frac{1 + 6 \dot{c}^2}{2c^2}\right)(1+f_R)  + \frac{1}{2}y^2(f-Rf_R) - \frac{y^2f_R''y' + (2yy'^2-y^2y'')f_R' - 6c\dot{c}\dot{f_R}y'^3}{2c^2y'^3} 
\end{equation}

\begin{equation}
    \label{7}
    p_r 2c(t)^2y'^2 =  \frac{y'^2}{y^2}\left(1-2\dot{c}^2-4c\ddot{c}\right)(1+f_R) - \frac{4c\dot{c}\dot{f_R}y'^2  + 2c^2\ddot{f_R}y'^2 - 3y f_R' y'}{y^2} - c^2y'^2(f-Rf_R)
\end{equation}

\begin{equation}
    \label{8}
     p_t c(t)^2y(r)^2 = \left(\frac{1}{2} - 2c\ddot{c} - \dot{c}^2\right)(1+f_R) - \frac{1}{2}c^2y^2(f-Rf_R) - \frac{4c\dot{c}\dot{f_R}y'^3 + 2c^2\ddot{f_R}y'^3 - y^2 f_R'' y' - (2yy'^2 - y^2y'')f_R'}{2y'^3} 
\end{equation}

\begin{equation}
    \label{9}
    -q \sqrt{2}c(t)y(r)y' =  \left(\frac{2 \dot{c} y'}{c y}\right)(1+f_R) - \frac{cy\dot{f_R}' - y\dot{c}f_R' - c\dot{f_R}y'}{cy}
\end{equation}

Where overhead dot indicates differentiation with respect to time coordinate $t$ and prime indicates differentiation with respect to radial coordinate $r$. Thus, equations (\ref{6}) to (\ref{9}) represents the field equations of gravitationally collapsing spherical stars in $ f(R) $ gravity govern by the interior metric (\ref{5}). 

\section{Junction conditions for gravitational collapse in $ f(R) $ gravity}
\label{sec3}
While solving field equations (\ref{6}) to (\ref{9}), we have to consider the junction conditions so that interior metric (\ref{5}) and exterior space-time smoothly match at the boundary of traditionally collapsing compact stars. Junction conditions also imply the continuity of exterior derivatives and Ricci scalar and its derivatives across the matching hypersurface. Thus, in this section, we derive the junction conditions for gravitationally collapsing stars in $ f(R) $ gravity governed by interior metric (\ref{5}), denoting interior metric by the $ ds^2_- $,

\begin{equation}
\label{10}
    ds^2_- =  - y(r)^2dt^2 + 2c(t)^2\left(\frac{dy}{dr}\right)^2dr^2 + c(t)^2y(r)^2(d\theta^2 + (sin^2\theta) d\phi^2)
\end{equation}

Now we denote the space-time interval at the boundary of the star as $ ds^2_{\Sigma} $. Thus, in co-moving coordinates, we write the $ ds^2_{\Sigma} $ as follows

\begin{equation}
\label{11}
    ds^2_{\Sigma} = g_{ij}d\xi^i d\xi^j = -d\tau^2 + \mathcal{R}^2(\tau)(d\theta^2 + (sin^2\theta) d\phi^2)
\end{equation}

Where $ \Sigma $ accounts for the hypersurface dividing interior and exterior space-time. Co-moving coordinates $ \xi^i = \tau$, $\theta$ and $ \phi $ describes the hypersurface $ \Sigma $. We consider the exterior space-time governed by the Vaidya metric as follows

\begin{equation}
\label{12}
    ds^2_+ = -\left( 1 - \frac{2m(v)}{\Tilde{r}} \right)dv^2 - 2dv d\Tilde{r} + \Tilde{r}^2(d\theta^2 + (sin^2\theta)d\phi^2)
\end{equation}

The first junction condition implies that metrics describing interior, boundary and exterior space-time must be continuous at matching hypersurface,

\begin{equation}
\label{13}
    ds^2_- =  ds^2_{\Sigma} = ds^2_+
\end{equation}

Now, the boundary equation for the interior metric (\ref{10}) is given by
\begin{equation}
\label{14}
    f(t,r) = r - r_{\Sigma} = 0
\end{equation}
Where $r_{\Sigma}$ is constant. Thus, the normal vector to that hypersurface is given by
\begin{equation}
\label{15}
    n_{\alpha}^- = \{ 0, \sqrt{2}c(t)\left(\frac{dy}{dr}\right)_{\Sigma}, 0, 0 \}
\end{equation} 
Putting $ dr = 0 $ in equation (\ref{10}) and comparing components of $ ds_-^2 $ and $ ds_{\Sigma}^2 $ at $ \Sigma $, we get

\begin{equation}
\label{16}
    y(r_{\Sigma})\dot{t} = 1
\end{equation}

\begin{equation}
\label{17}
    c(t)y(r_{\Sigma}) = R(\tau)
\end{equation}

Where, the dot stands for $ \frac{d}{d \tau}. $Now, boundary equation for the exterior metric (\ref{12}) is given by

\begin{equation}
    \label{18}
    f(v,\Tilde{r}) = \Tilde{r} - \Tilde{r}_{\Sigma}(v) = 0
\end{equation}
Which yields the following normal vector to the hypersurface (\ref{18})

\begin{equation}
\label{19}
    n_{\alpha}^+ = \left(2\frac{d\Tilde{r}_{\Sigma}(v)}{dv} + 1 - \frac{2m}{\Tilde{r}_{\Sigma}} \right)^{-\frac{1}{2}}\left( -\frac{d\Tilde{r}_{\Sigma}(v)}{dv}, 1, 0, 0 \right)
\end{equation}

The junction condition (\ref{11}) for (\ref{11}) and (\ref{13}) leads to the conditions

\begin{equation}
\label{20}
    \Tilde{r}_{\Sigma}(v) = \mathcal{R}(\tau)
\end{equation}

\begin{equation}
\label{21}
   \left(2\frac{d\Tilde{r}(v)}{dv} + 1 - \frac{2m}{\Tilde{r}} \right)_{\Sigma} = \left( \frac{1}{\dot{v}^2} \right)_{\Sigma} 
\end{equation}

The second junction condition implies that extrinsic curvature $ K_{ij} $ must be continuous across the hypersurface $ \Sigma $.

\begin{equation}
\label{22}
    [K_{ij}] = K_{ij}^+ - K_{ij}^- = 0
\end{equation}

Where the extrinsic curvature is given by

\begin{equation}
    \label{23}
    K^{\pm}_{ij} = -n_{\alpha}^{\pm} \frac{\partial^2 x^{\alpha}_{\pm}}{\partial \xi^i \partial \xi^j} - n_{\alpha}^{\pm} \Gamma^{\alpha}_{\mu \nu} \frac{\partial x^{\mu}_{\pm}}{\partial \xi^i} \frac{\partial x^{\nu}_{\pm}}{\partial \xi^j}
\end{equation}

Now this junction condition has been developed for the following general time-dependant metric by many authors
\begin{equation}
\label{24}
    ds_-^2 = -e^{2\nu(t,r)}dt^2 + e^{2\psi(t,r)}dr^2 + Q^2(t,r)(d\theta^2 + sin^2\theta d\phi^2)
\end{equation}

Junction condition (\ref{22}) for the interior metric (\ref{24}) and exterior metric (\ref{12}) is found to be

\begin{equation}
\label{25}
    M(t,r) = \frac{Q}{2}\left[ 1 - e^{-2\psi}\left( \frac{dQ}{dr} \right)^2 +  e^{-2\nu}\left( \frac{dQ}{dt} \right)^2  \right]
\end{equation}

\begin{equation}
\label{26}
    \frac{Q}{2}e^{-(\nu+\psi)}\left( 2\frac{\dot{Q}'}{Q} - 2\frac{\dot{Q}}{Q}\frac{\dot{\psi}}{\psi} - 2\frac{\nu'}{\nu}\frac{\dot{Q}}{Q}\right) +
    \frac{Q}{2} e^{-2\nu}\left( 2\frac{\ddot{Q}}{Q} - 2\frac{\dot{Q}}{Q}\frac{\dot{\nu}}{\nu} + \frac{e^{2\nu}}{Q^2} + \frac{\dot{Q}^2}{Q^2} - e^{2(\nu-\psi)}\left( \frac{Q'^2}{Q^2} - 2\frac{\nu'}{\nu}\frac{Q'}{Q} \right) \right) \bigg |_{\Sigma} = 0
\end{equation}

By comparing the interior metric of our case (\ref{10}) with generalized interior metric, (\ref{24}) we can derive the junction condition corresponding to equation (\ref{22}) for the metric (\ref{10}) as follow

\begin{equation}
\label{27}
 M(t,r) =  \frac{c y}{2}\left( \frac{1}{2} + \dot{c}^2 \right)
\end{equation}

\begin{equation}
\label{28}
\frac{1}{2\sqrt{2}y'}\left(\frac{2\dot{c}y'}{c y} - 2\frac{\dot{c}^2}{c^2}\frac{1}{log(\sqrt{2}c y')} - \frac{2\dot{c}}{c y log(y)} + \frac{2\sqrt{2}\ddot{c}y'}{y} +  \frac{\sqrt{2}y'}{c y} + \frac{\sqrt{2}y'\dot{c}^2}{c y} - \frac{y'}{\sqrt{2}c y} + \frac{\sqrt{2}}{c y log(y)} \right)\bigg |_{\Sigma} = 0    
\end{equation}
Where $M(t,r)$ denotes the Misner-Sharp mass function equals to the Schwarzschild mass when evaluated at the boundary $ \Sigma $. 
\par Final two junction conditions implies that the Ricci scalar and its derivative must be continuous across the hypersurface $ \Sigma $,

\begin{equation}
\label{29}
    [R]  = 0 \quad\mathrm{and}\quad n^{\alpha}[\partial_{\alpha}R] = 0
\end{equation}

where Ricci scalar for the interior metric (\ref{10}) is given by 

\begin{equation}
\label{30}
     R = \frac{6\dot{c}^2 + 6c\ddot{c} - 1}{c^2y^2}
\end{equation}

\section{Gravitational collapse model in $ f(R) = k R^m $ gravity}
\label{sec4}
Equations (\ref{6}) to (\ref{9}) shows that there are six unknowns ($ \rho $, $p_r$, $p_t$, $q$, $c$ and $ y $) but only four equations. Thus we can choose any two unknowns to solve field equations (\ref{6}) to (\ref{9}). Here we will choose the forms of $ y(r) $ and $ c(t) $ such that model obeys the junction conditions presented in the previous section. It is straightforward to see that the following metric form can obey all junction conditions, including the continuity of the Ricci scalar and its derivatives across the boundary

\begin{equation}
\label{31}
    y(r) = \left(1 - \frac{r}{a}\right)^{-n} 
\end{equation}
Where $ a $ denotes the constant value of radius of the collapsing star in co-moving coordinates and $n$ is the constant having condition $ n \ge 1 $. Now we choose the form of $ c(t) $ such that model simplifies considerably such that $ 6\dot{c}^2 + 6c\ddot{c} = 0 $ holds. The condition gives the following form of $ c(t) $,

\begin{equation}
\label{32}
    c(t) = \sqrt{1 - b t}
\end{equation}

where $ b $ is the integration constant. Putting the values of the $ y(r) $ and $ c(t) $ from equations (\ref{31}) and (\ref{32}) in equations (\ref{6}) to (\ref{9}), we calculate the exact solutions describing physical quantities like energy density, two pressures and heat flux density of the gravitationally collapsing stars. Here we choose $ n = 2 $ in equation (\ref{31}) and use general form of $ c(t) $ as given in (\ref{32}). Thus, we solve equations (\ref{6}) to (\ref{9}) for the following physically reliable forms of $ y(r) $ and $ c(t) $,

\begin{equation}
    \label{33}
    y(r) = \left(1 - \frac{r}{a}\right)^{-2} \quad\mathrm{and}\quad  c(t) = \sqrt{1 - b t}
\end{equation}

From equation (\ref{30}), we calculate Ricci scalar $ R $ for the metric components (\ref{33}) as follow

\begin{equation}
    \label{34}
    R = -\frac{(1 - \frac{r}{a})^4}{1-b t}
\end{equation}

Here we consider the model $ R + f(R) = R + k R^m  $, which gives the identities $ f(R) = k R^m $ and $ f_R  = k m R^{m-1} $. Putting all these values in equations (\ref{6}) to (\ref{9}), we get the following values of energy density, two pressures and heat flux density for gravitationally collapsing stars 

\begin{multline}
    \label{35}
    \rho(t,r) = \frac{(1 - \frac{r}{a})^4}{4(1-b t)^{m+1}}[(2-2b t+3b^2)(1-b t)^{m-1} \\+ k(-1)^{m+1}\left(1-\frac{r}{a}\right)^{4m-4}\{ (-8m^3 + 20m^2 -8m -2)(1-b t) + b^2(9m - 6m^2) \}]
\end{multline}

\begin{multline}
    \label{36}
    p_r(t,r) = \frac{(1 - \frac{r}{a})^4}{4(1-b t)^{m+1}}[(2-2b t+b^2)(1-b t)^{m-1} \\+ k(-1)^{m+1}\left(1-\frac{r}{a}\right)^{4m-4}\{ (-12m^2 + 12m + 2)(1-b t) + (-4b^2m^3 + 4b(b+1)m^2 - b(4-b)m) \}]
\end{multline}

\begin{multline}
    \label{37}
    p_t(t,r) = \frac{(1 - \frac{r}{a})^4}{4(1-b t)^{m+1}}[(2-2b t+b^2)(1-b t)^{m-1} \\+ k(-1)^{m+1}\left(1-\frac{r}{a}\right)^{4m-4}\{ (8m^3 - 20m^2 + 12m + 2)(1-b t) - m b^2(4(m-1)^2 - 1) \}]
\end{multline}

\begin{equation}
    \label{38}
    q(t,r) = \frac{(1 - \frac{r}{a})^4}{\sqrt{2}(1-b t)^{m+\frac{1}{2}}}[b(1-b t)^{m-1} + k(-1)^{m+1}\left(1-\frac{r}{a}\right)^{4m-4}\{b m - 2m^2b(m - 1) \}]
\end{equation}
    
Equations (\ref{35}) to (\ref{38}) are analytical solutions describing physical quantities of gravitationally collapsing stars in modified $ f(R) = k R^m $ gravity.

\section{Energy conditions for gravitationally collapsing stars}
\label{sec5}
To describe the realistic compact stars undergoing gravitational collapse, the model of gravitational collapse must obey certain energy conditions. Thus, in this section we describe the energy conditions to be applied so that model become physically acceptable. We consider the shear-less matter distribution, which is essential the case for separable form of interior space-time metric (\ref{5}). Thus, we apply the following energy conditions on the model.

\subsection{Eigenvalues of energy-momentum tensor must be real}
\label{seca}
The eigenvalues of the energy-momentum tensor must be real for a physically acceptable model of any compact star. The eigenvalues of the energy-momentum tensor can be found by equation $ |T_{\alpha \beta} - \lambda g_{\alpha \beta}| = 0 $. The detailed calculation performed is in the reference \cite{2008GReGr..40.2149P}. From that it can be derived that for shear-less matter distribution, these conditions can be written as

\begin{equation}
    \label{39}
    |\rho + p_r| - 2|q| \ge 0
\end{equation}

\begin{equation}
    \label{40}
    \rho - p_r + 2p_t + \Delta \ge 0
\end{equation}
\\
where, $\Delta = \sqrt{(\rho + p_r)^2 - 4q^2} $.

\subsection{Weak energy condition}
\label{secb}
The weak energy condition must be obeyed for a physically acceptable model of gravitational collapse. The weak energy condition states that if $ \lambda_0 $ signifies the eigenvalue corresponding to the time-like eigenvector, then $ -\lambda_0 \ge 0 $ must be true. As a result, the shear-less fluid's weak energy condition becomes

\begin{equation}
\label{41}
    \rho - p_r + \Delta \ge 0
\end{equation}

\subsection{Dominant energy conditions}
\label{secc}
Any physically acceptable model must also obey the dominant energy conditions. If $ \lambda_i $ signifies the eigenvalues corresponding to the space-like eigenvectors, then it must follow the condition given by $ \lambda_0 \le \lambda_i \le \lambda_0 $. Thus, the dominant energy conditions for shear-less fluid can be written in terms of physical quantities of the star as follow

\begin{equation}
\label{42}
    \rho - p_r - 2p_t + \Delta \ge 0
\end{equation}

\begin{equation}
\label{43}
    \rho - p_r \ge 0
\end{equation}

\subsection{Strong energy condition}
\label{secd}
By imposing the last strong energy condition, a physically acceptable model can be found. The presence of a strong energy condition indicates that $ \lambda_0 + \sum_{i} \lambda_i \ge 0 $. Which implies that physical quantities of the star mast obey following condition

\begin{equation}
\label{44}
    2p_t + \Delta \ge 0
\end{equation}

\par Thus, physically acceptable model of stars undergoing gravitational collapse can be found by imposing all energy conditions from (\ref{39}) to (\ref{44}).

\section{Physical analysis of the model}
\label{sec6}
In this section we present physical analysis for the different values of $ k $ and $ m $. Equations (\ref{35}) to (\ref{38}) represents analytical solutions for general case of $ R + f(R) = R + k R^m $. Here we choose different values of $ k $ and $ m $ and apply energy conditions for each case to generate physically acceptable models of gravitational collapse. 

\subsection{Physical analysis with $ k = 0 $}
Firstly, we choose the value of $ k = 0 $. So that the field equations becomes essentially of General Relativity. Thus, with $ k = 0 $, equations (\ref{35}) to (\ref{38}) describes gravitational collapse in General Relativity. By putting $ k = 0 $ in them we get the following analytical forms of physical quantities by choosing $ a = b = 1 $

\begin{equation}
    \label{45}
    \rho(t,r) = \frac{(1 - r)^4}{4(1- t)^{2}}(5-2 t)
\end{equation}

\begin{equation}
    \label{46}
    p_r(t,r) = \frac{(1 - r)^4}{4(1- t)^{2}}(3-2 t)
\end{equation}

\begin{equation}
    \label{47}
    p_t(t,r) = \frac{(1 - r)^4}{4(1- t)^{2}}(3-2 t)
\end{equation}

\begin{equation}
    \label{48}
    q(t,r) = \frac{(1 - r)^4}{\sqrt{2}(1- t)^{\frac{3}{2}}}
\end{equation}

\begin{figure}[htp]
\centering
\includegraphics[width=.33\textwidth]{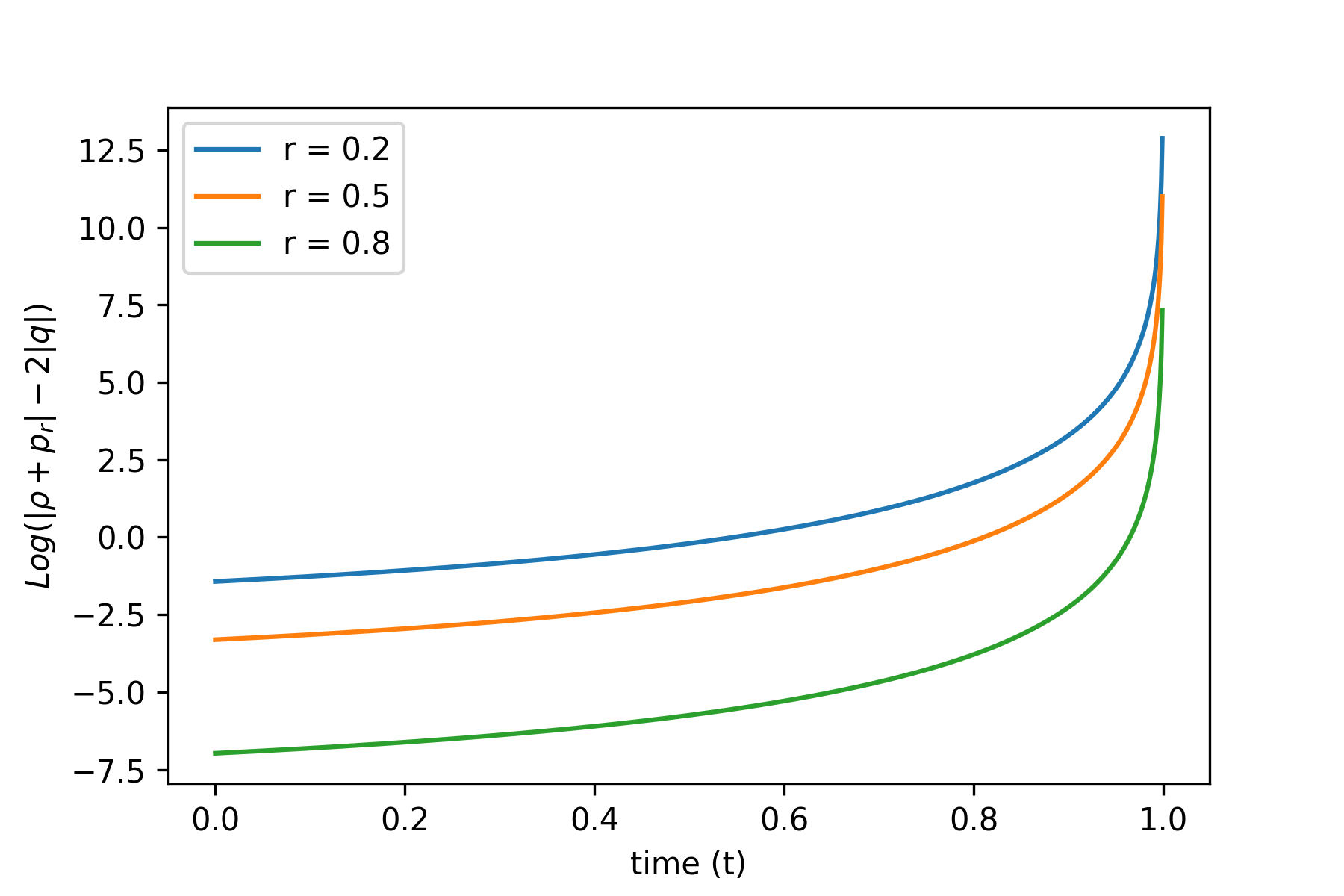}\hfill
\includegraphics[width=.33\textwidth]{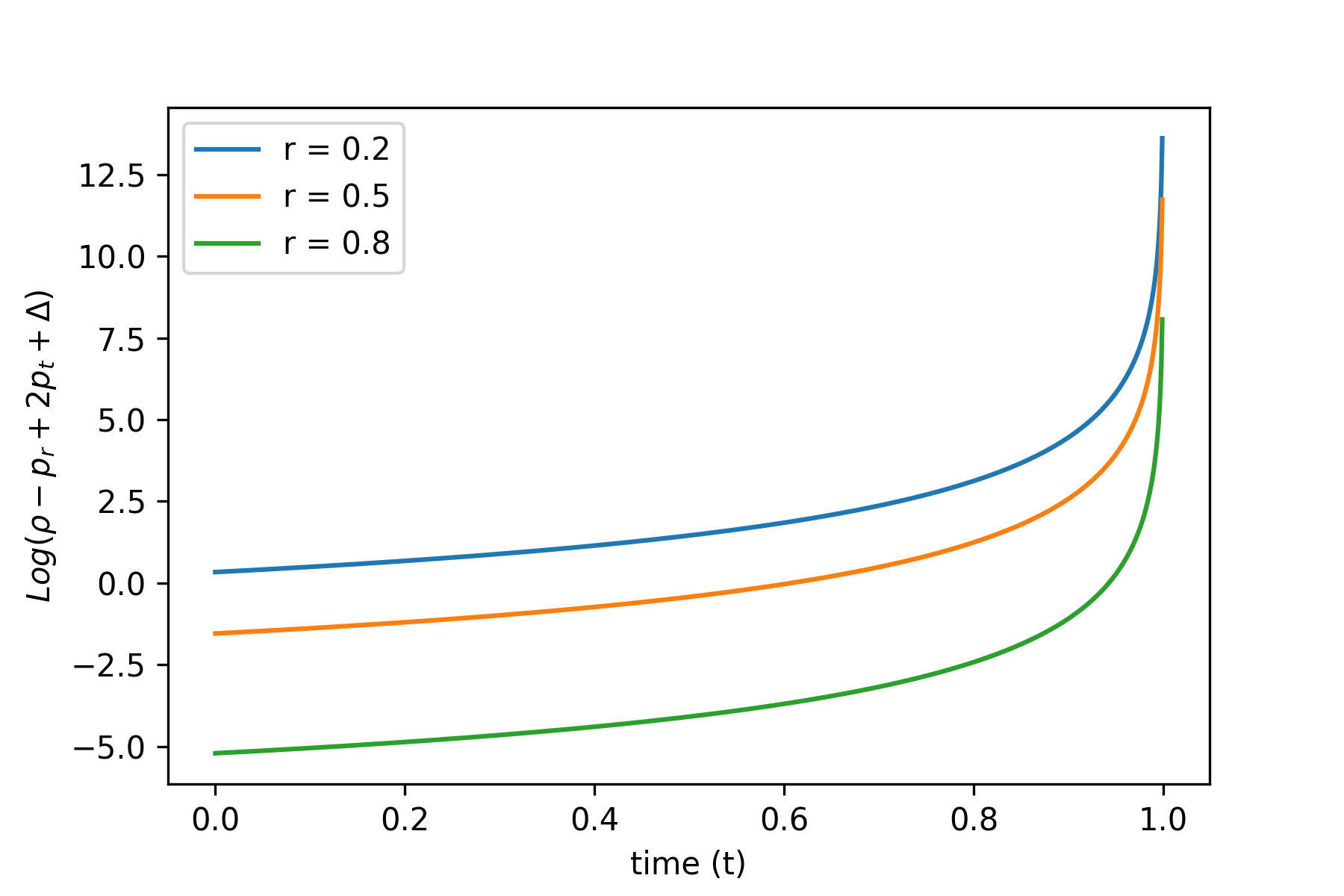}\hfill
\includegraphics[width=.33\textwidth]{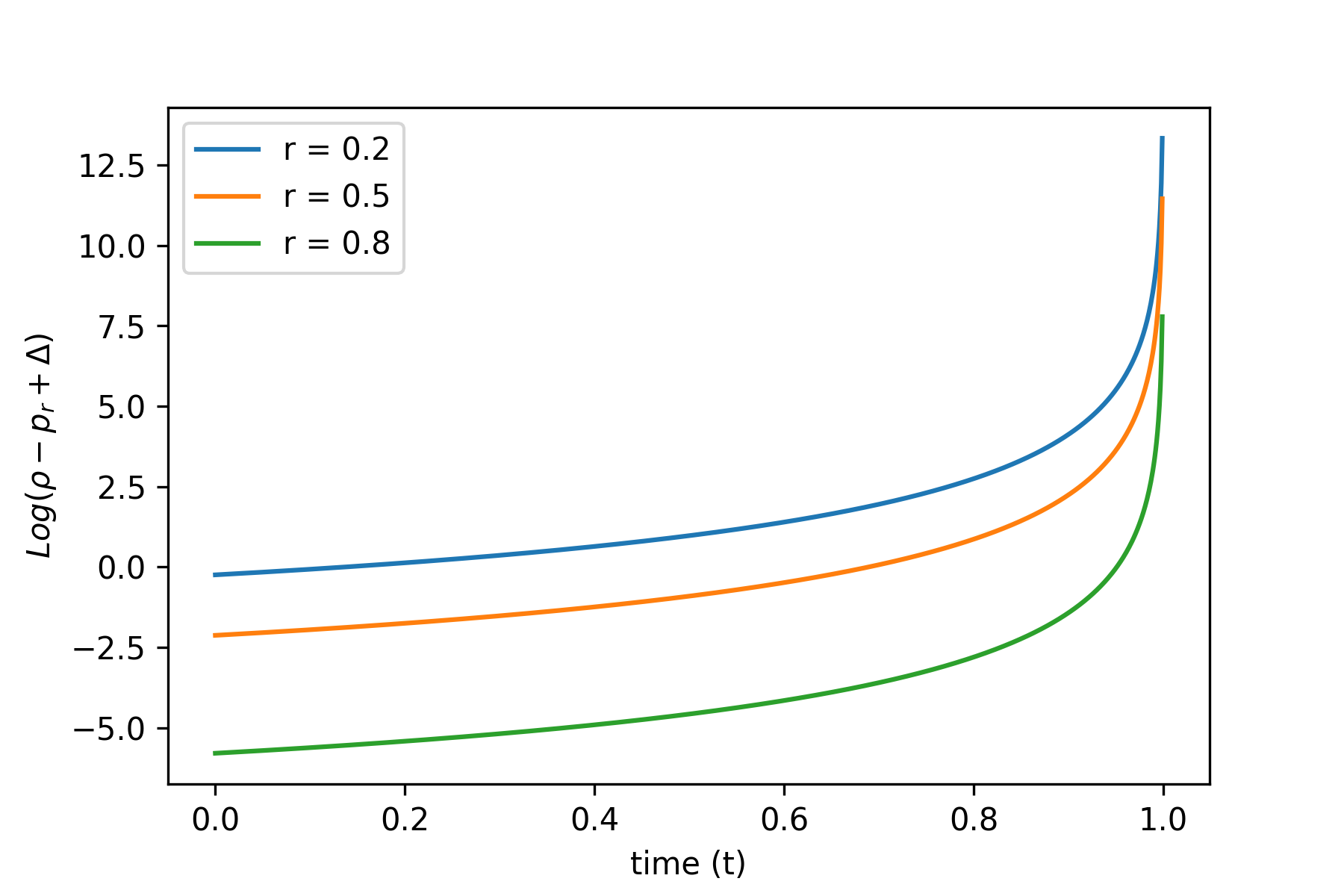}\hfill
\includegraphics[width=.33\textwidth]{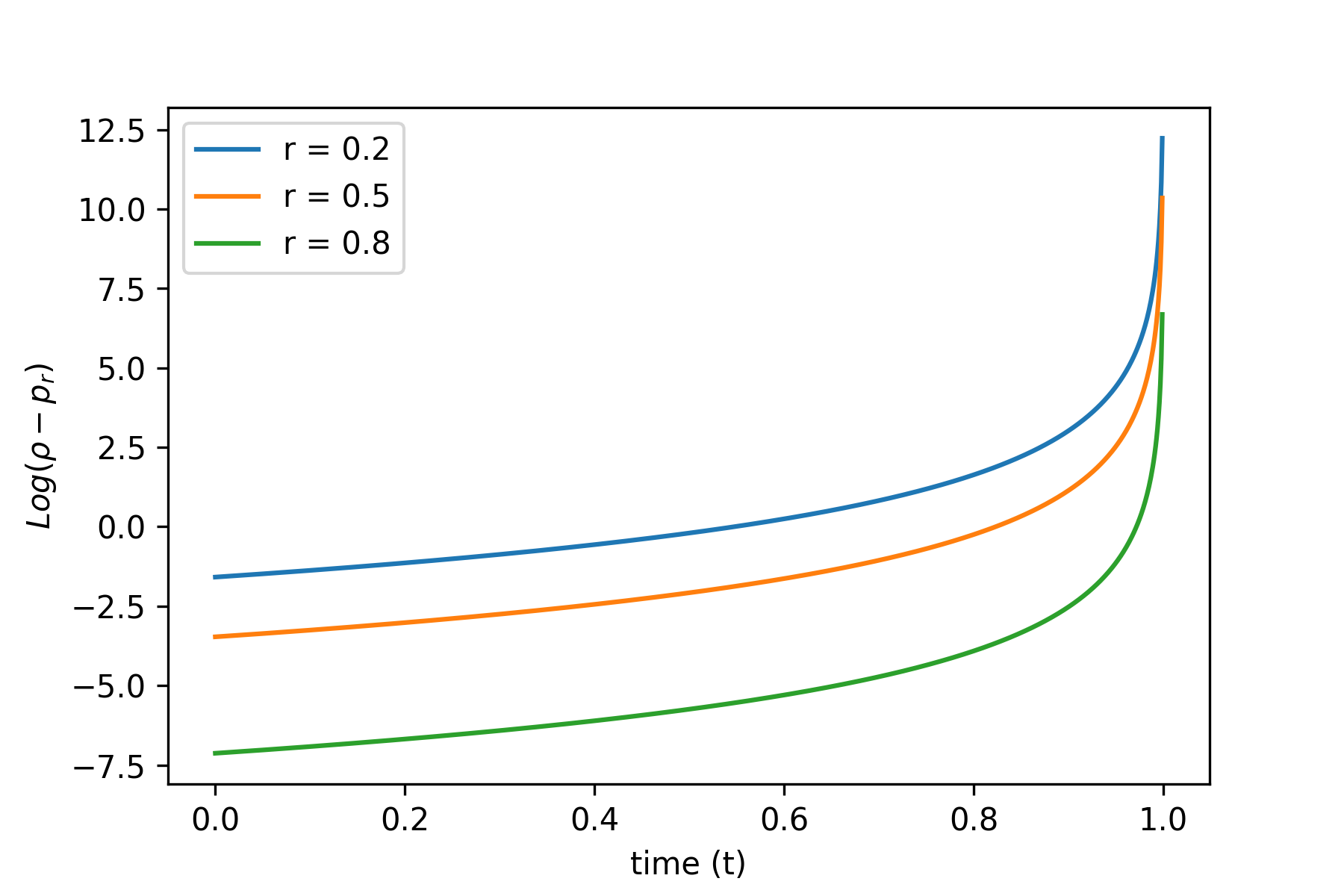}\hfill
\includegraphics[width=.33\textwidth]{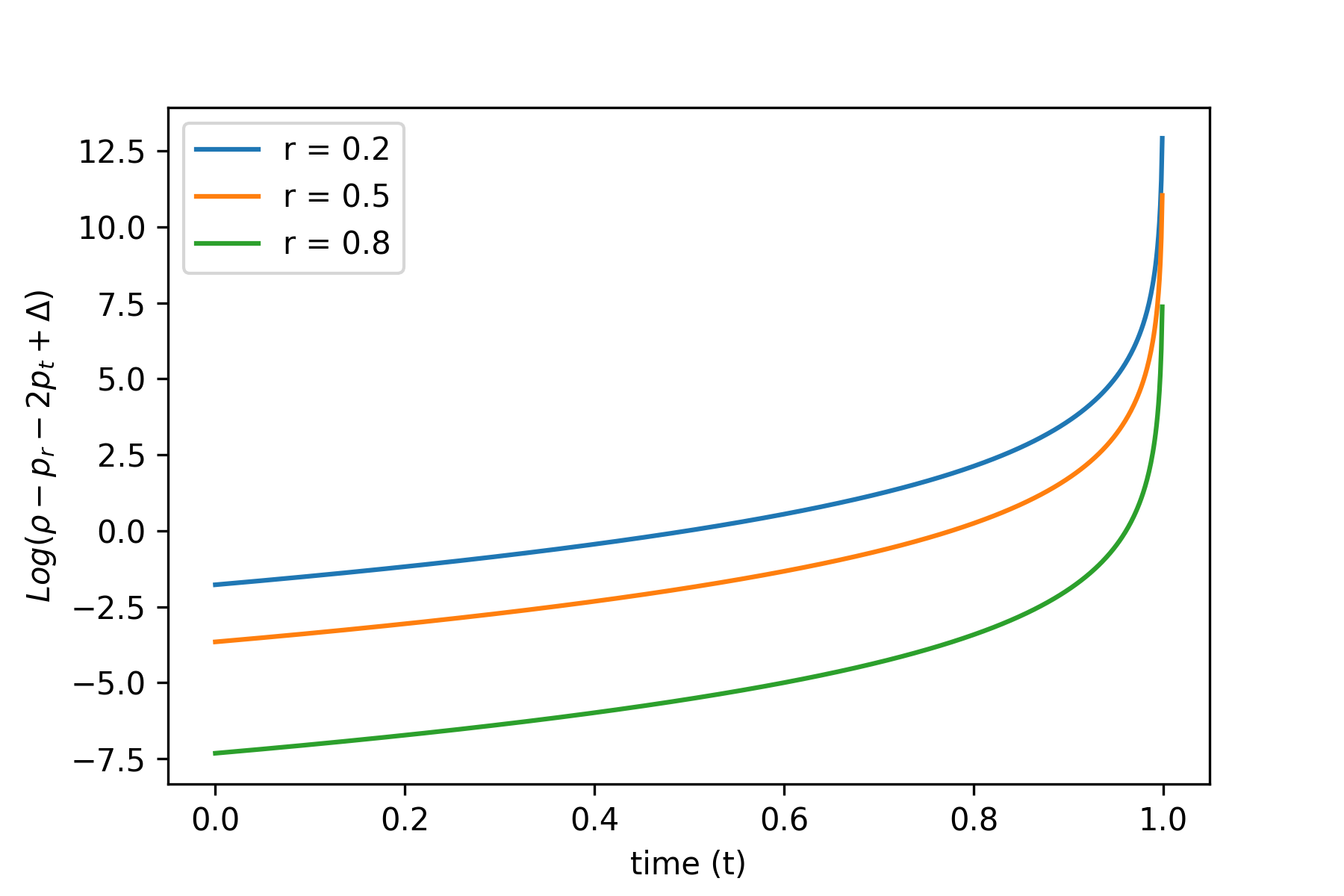}\hfill
\includegraphics[width=.33\textwidth]{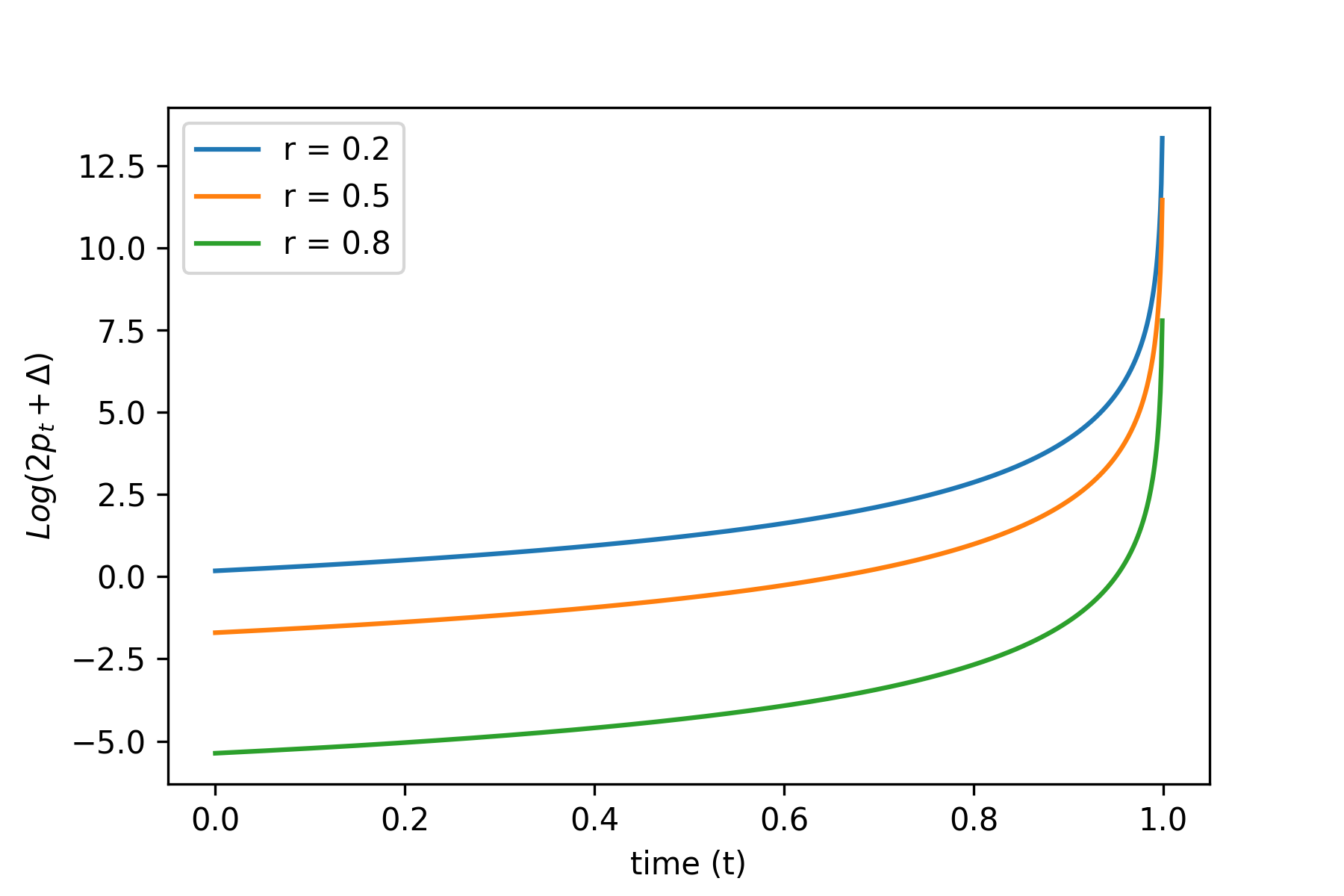}\hfill
\caption{Energy conditions for the model of gravitational collapse with $ k = 0 $}
\label{fig1}
\end{figure}

Equations (\ref{45}) to (\ref{48}) shows that gravitational collapse starts at $ t = 0 $ and ends up in singularity $ t = 1 $ with blowing up all physical quantities. It can be seen from the equations (\ref{46}) and (\ref{47}) that anisotropy of collapsing stars for $ k = 0 $ identically vanishes for all time at all interior points of the star. However, it is not the case for the non-zero value of $ k $, providing the model in modified $ f(R) $ gravity. Also, we plot all the energy conditions described in section \ref{sec5}. All the graphs are presented in FIG \ref{fig1} show that this model satisfies all the energy conditions, and thus it is physically acceptable. 

\subsection{Physical analysis of the model $ f(R) = k R^2 $}
Now we choose the value of $ m = 2 $. Thus, the model becomes $ R + f(R) = R + k R^2 $. We put the values of $ m = 2 $, $ k = 10^{-3} $, and $ a = b = 1 $ in equations (\ref{35}) to (\ref{38}) and thus, the equations becomes

\begin{equation}
    \label{49}
    \rho(t,r) = \frac{(1 - r)^4}{4(1- t)^{3}}[(5-2t)(1- t) + k\left(1-r\right)^{4}(8-2t)]
\end{equation}

\begin{equation}
    \label{50}
    p_r(t,r) = \frac{(1 - r)^4}{4(1- t)^{3}}[(3-2t)(1- t) + k\left(1-r\right)^{4}(28-22t)]
\end{equation}

\begin{equation}
    \label{51}
    p_t(t,r) =  \frac{(1 - r)^4}{4(1- t)^{3}}[(3-2t)(1- t) - k\left(1-r\right)^{4}(4-10t)]
\end{equation}

\begin{equation}
    \label{52}
    q(t,r) = \frac{(1 - r)^4}{\sqrt{2}(1- t)^{\frac{5}{2}}}[1- t + 6k\left(1-r\right)^{4}]
\end{equation}

\begin{figure}[htp]
\centering
\includegraphics[width=.33\textwidth]{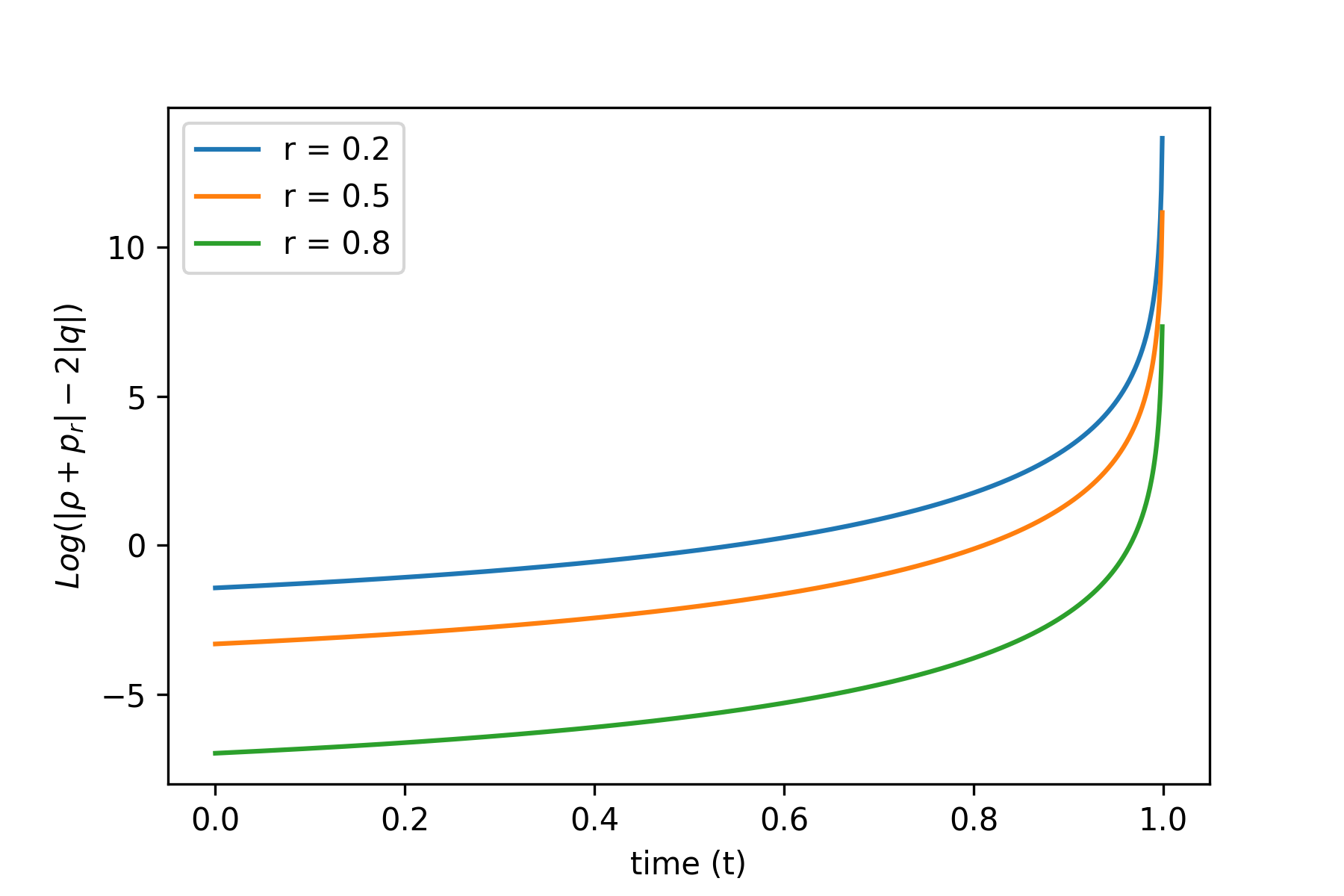}\hfill
\includegraphics[width=.33\textwidth]{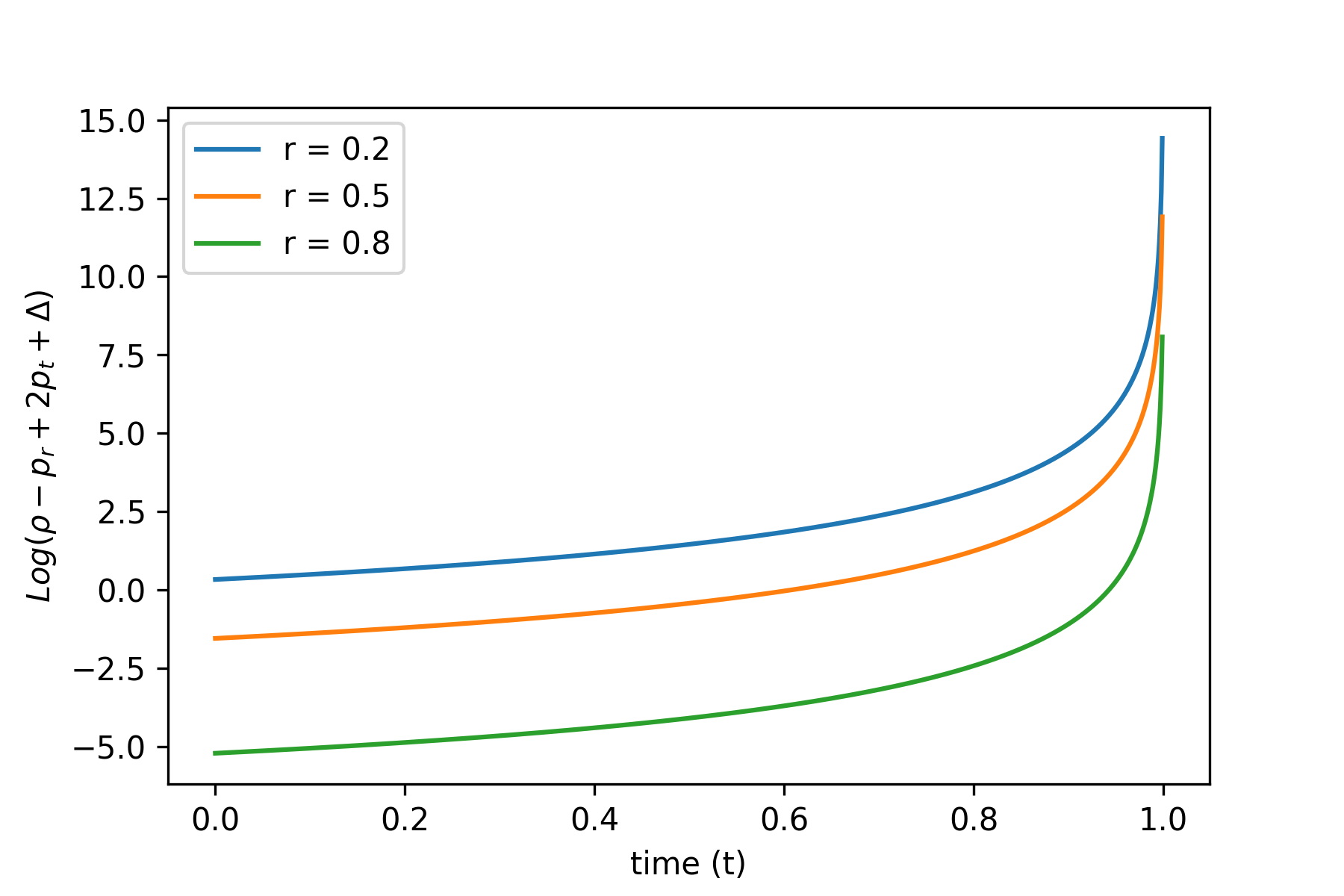}\hfill
\includegraphics[width=.33\textwidth]{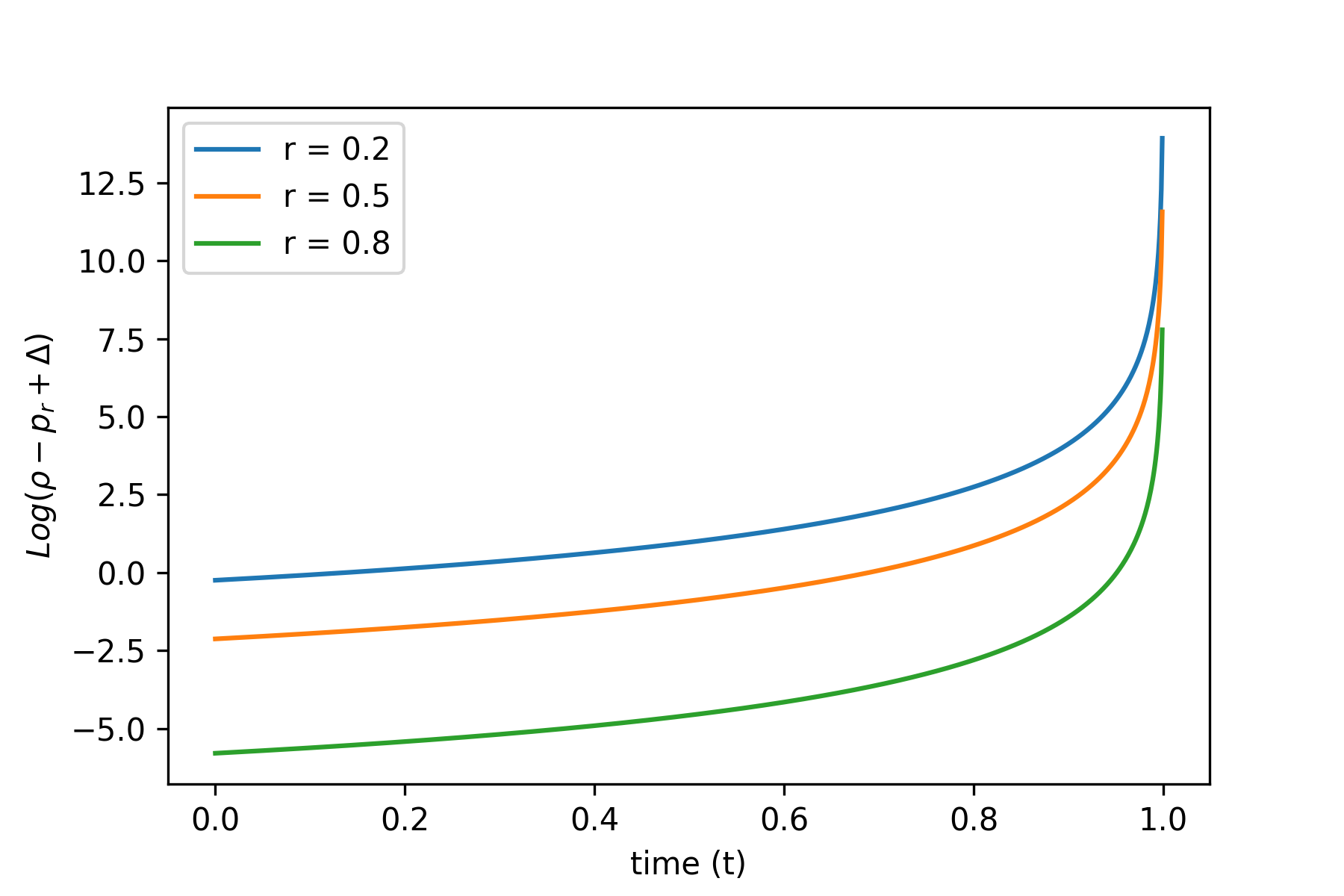}\hfill
\includegraphics[width=.33\textwidth]{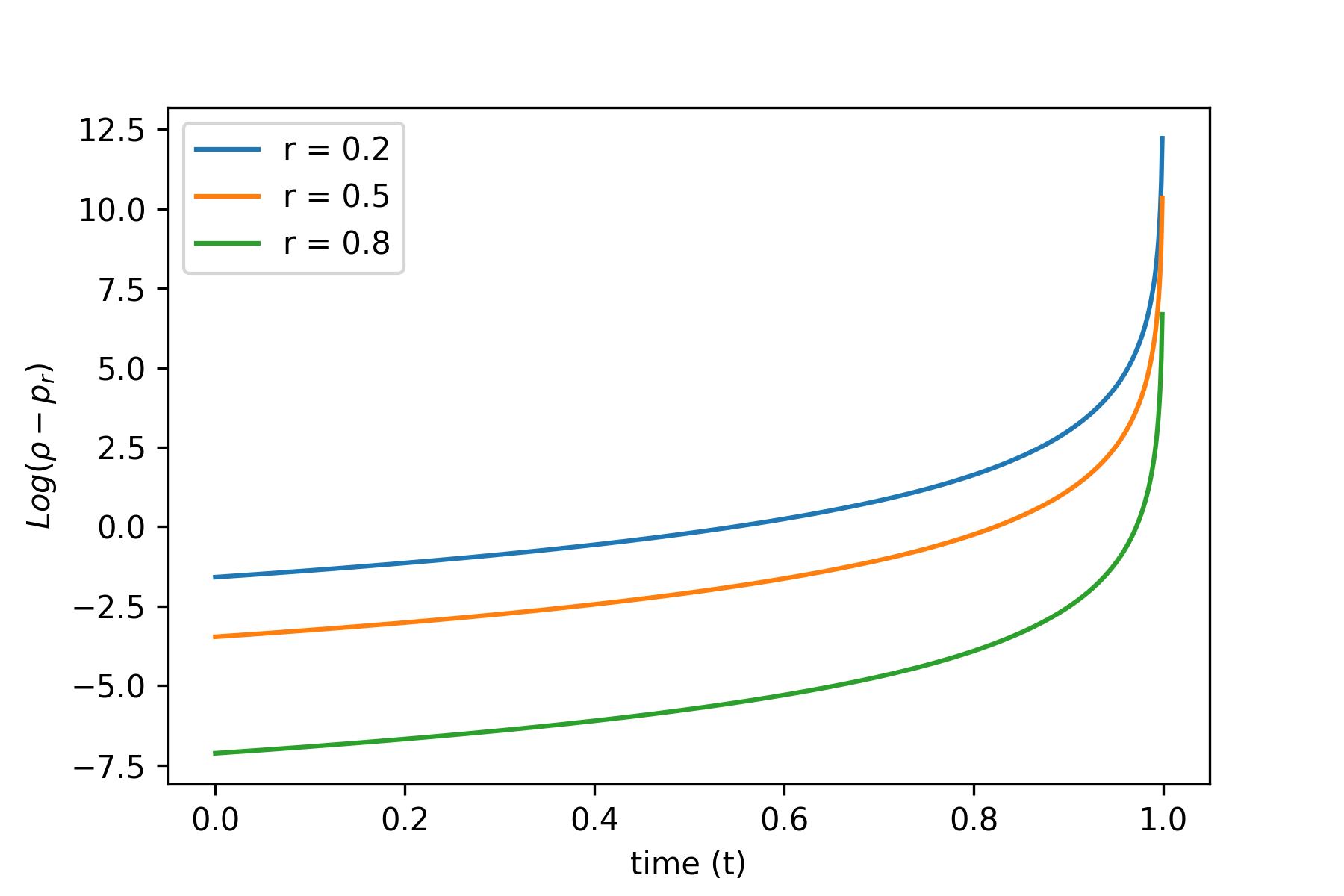}\hfill
\includegraphics[width=.33\textwidth]{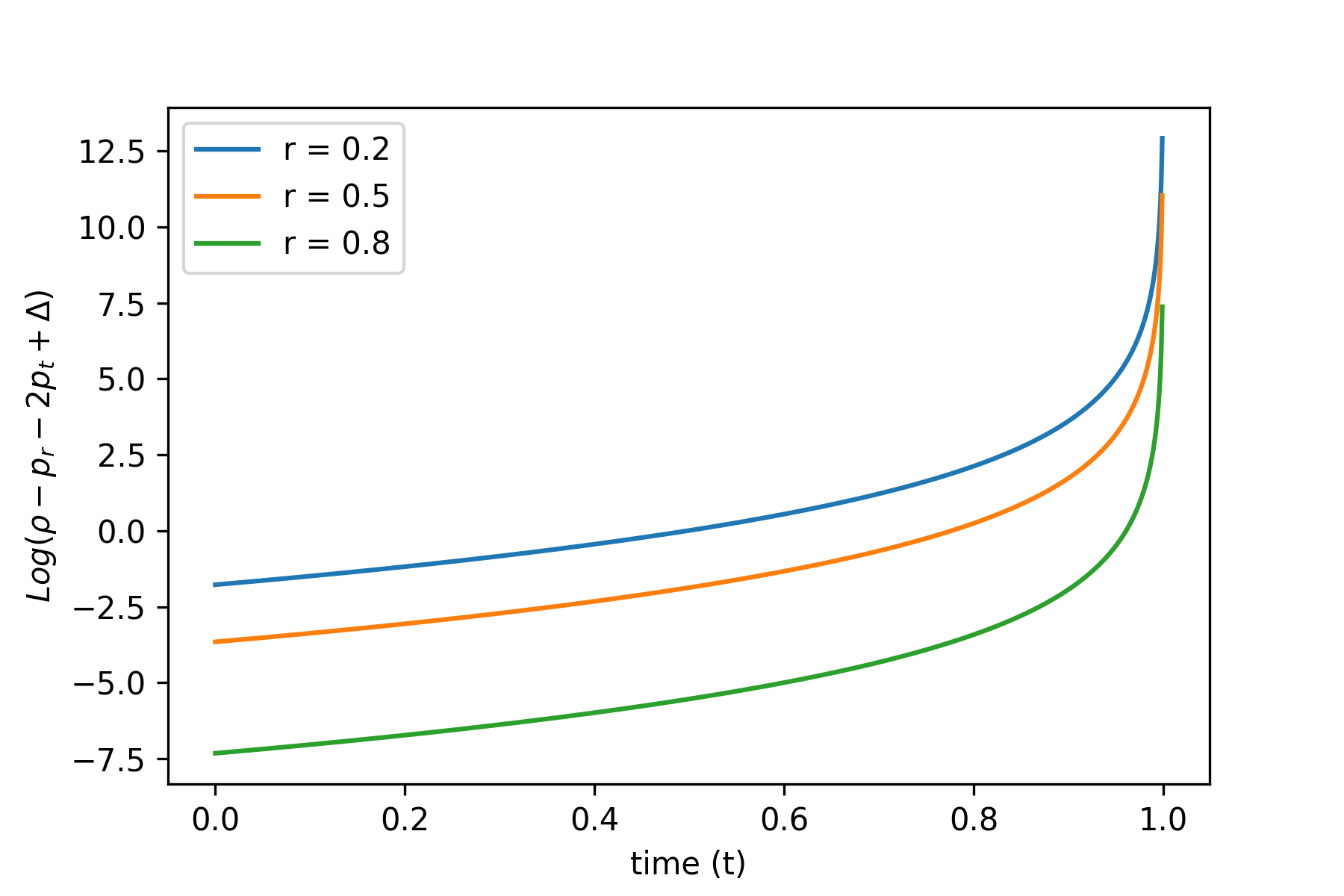}\hfill
\includegraphics[width=.33\textwidth]{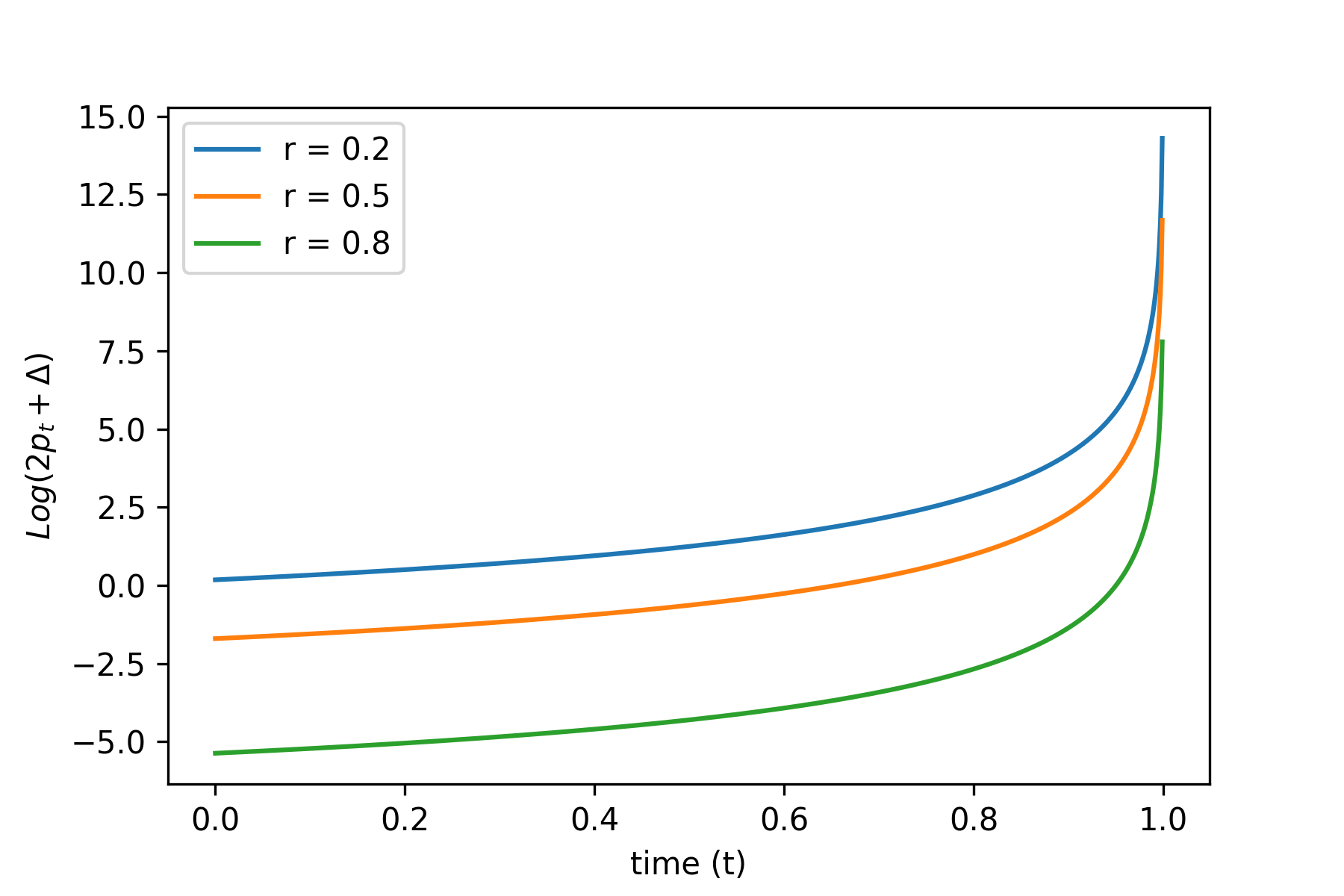}\hfill
\caption{Energy conditions for the model of gravitational collapse with $ f(R) = kR^2 $}
\label{fig2}
\end{figure}

Equations (\ref{49}) to (\ref{52}) represents analytical solutions of gravitational collapse in $ R^2 $ corrected $ f(R) $ gravity. Equations (\ref{49}) to (\ref{52}) shows that for this model also collapse starts at $ t = 0 $ and it ends up in singularity at $ t = 1 $, due to the choice of $ b = 1 $. Equations (\ref{50}) and (\ref{51}) shows that anisotropy doesn't vanish at the interior of star, unlike vanishing anisotropy in case of collapse governed by general relativity. We plot all the energy conditions for this model in FIG \ref{fig2}. All the graphs presented in FIG \ref{fig2}. shows that this model satisfy all the energy conditions and thus it is physically acceptable. 

\subsection{Physical analysis of the model $ f(R) = k R^4 $}
Finally we choose $ R + f(R) = R + k R^4 $ model to describe gravitational collapse of compact stars. 
We put the values of $ m = 4 $, $ k = 10^{-7} $, and $ a = b = 1 $ in equations (\ref{35}) to (\ref{38}) and thus, the equations becomes

\begin{equation}
    \label{53}
    \rho(t,r) = \frac{(1 - r)^4}{4(1- t)^{5}}[(5-2t)(1- t)^3 + k\left(1-r\right)^{12}(286-226t)]
\end{equation}

\begin{equation}
    \label{54}
    p_r(t,r) = \frac{(1 - r)^4}{4(1- t)^{5}}[(3-2t)(1- t)^3 + k\left(1-r\right)^{12}(282-142t)]
\end{equation}

\begin{equation}
    \label{55}
    p_t(t,r) =  \frac{(1 - r)^4}{4(1- t)^{5}}[(3-2t)(1- t)^3 - k\left(1-r\right)^{12}(102-242t)]
\end{equation}

\begin{equation}
    \label{56}
    q(t,r) = \frac{(1 - r)^4}{\sqrt{2}(1- t)^{\frac{9}{2}}}[(1- t)^3 + 92k\left(1-r\right)^{12}]
\end{equation}

\begin{figure}[htp]
\centering
\includegraphics[width=.33\textwidth]{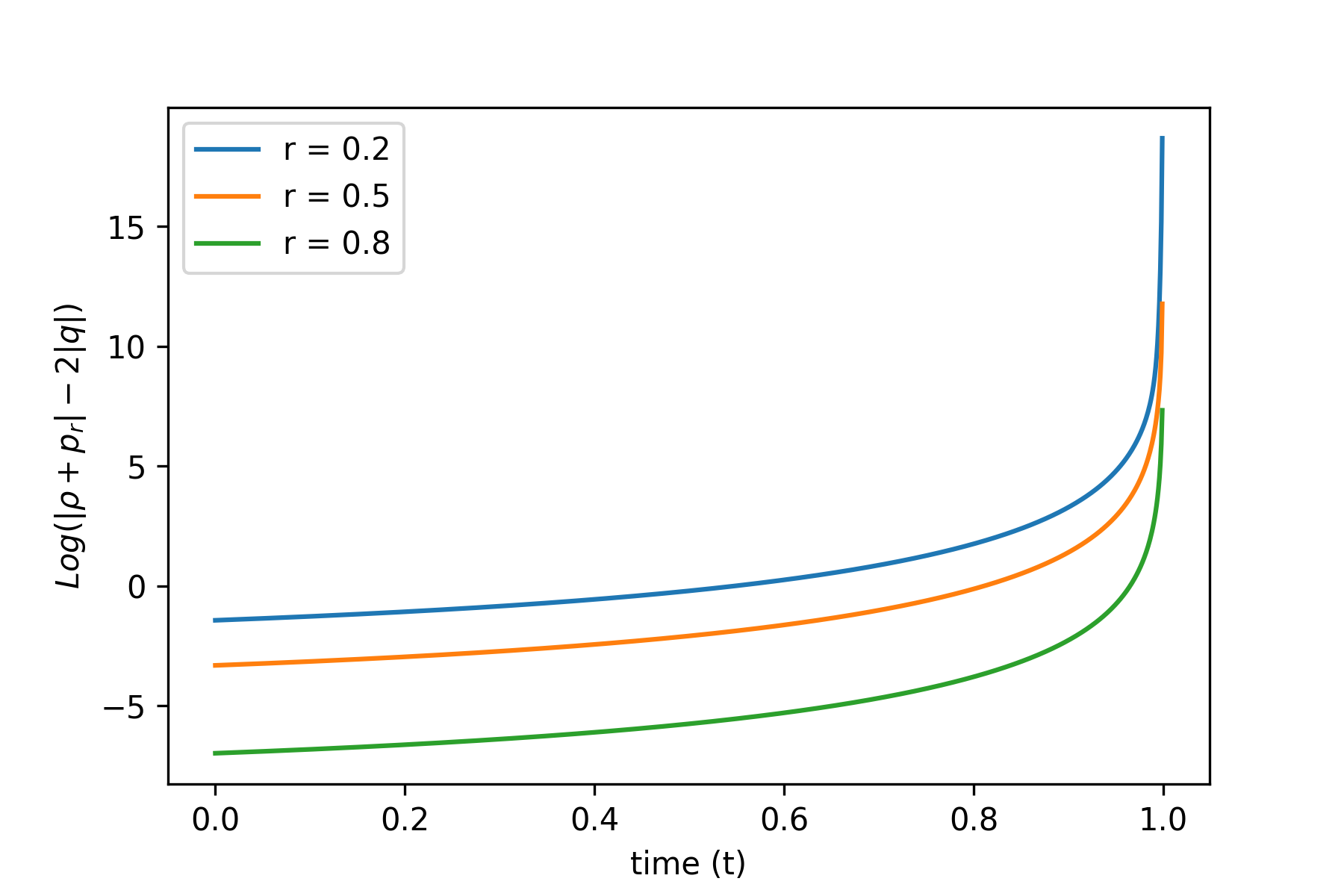}\hfill
\includegraphics[width=.33\textwidth]{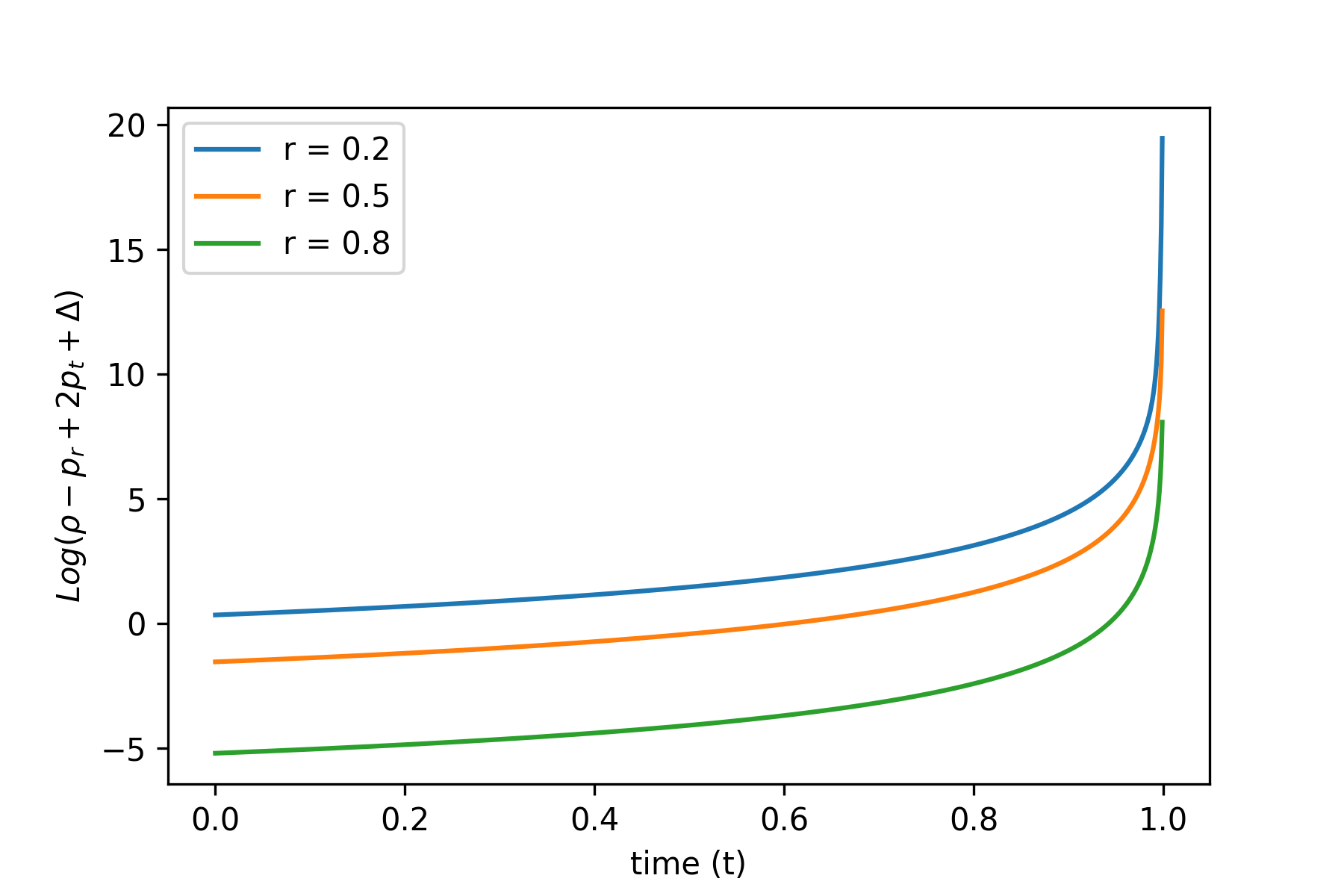}\hfill
\includegraphics[width=.33\textwidth]{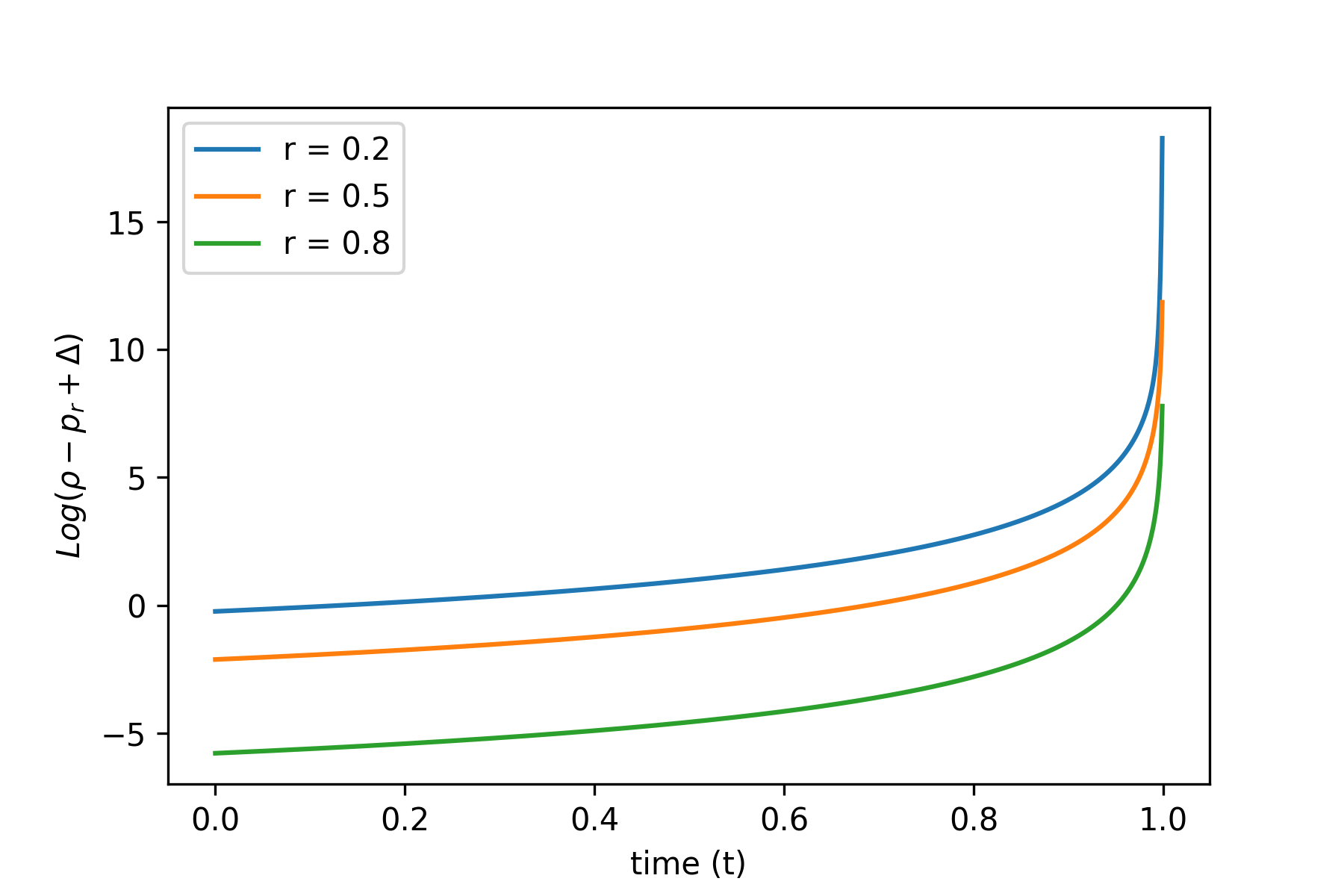}\hfill
\includegraphics[width=.33\textwidth]{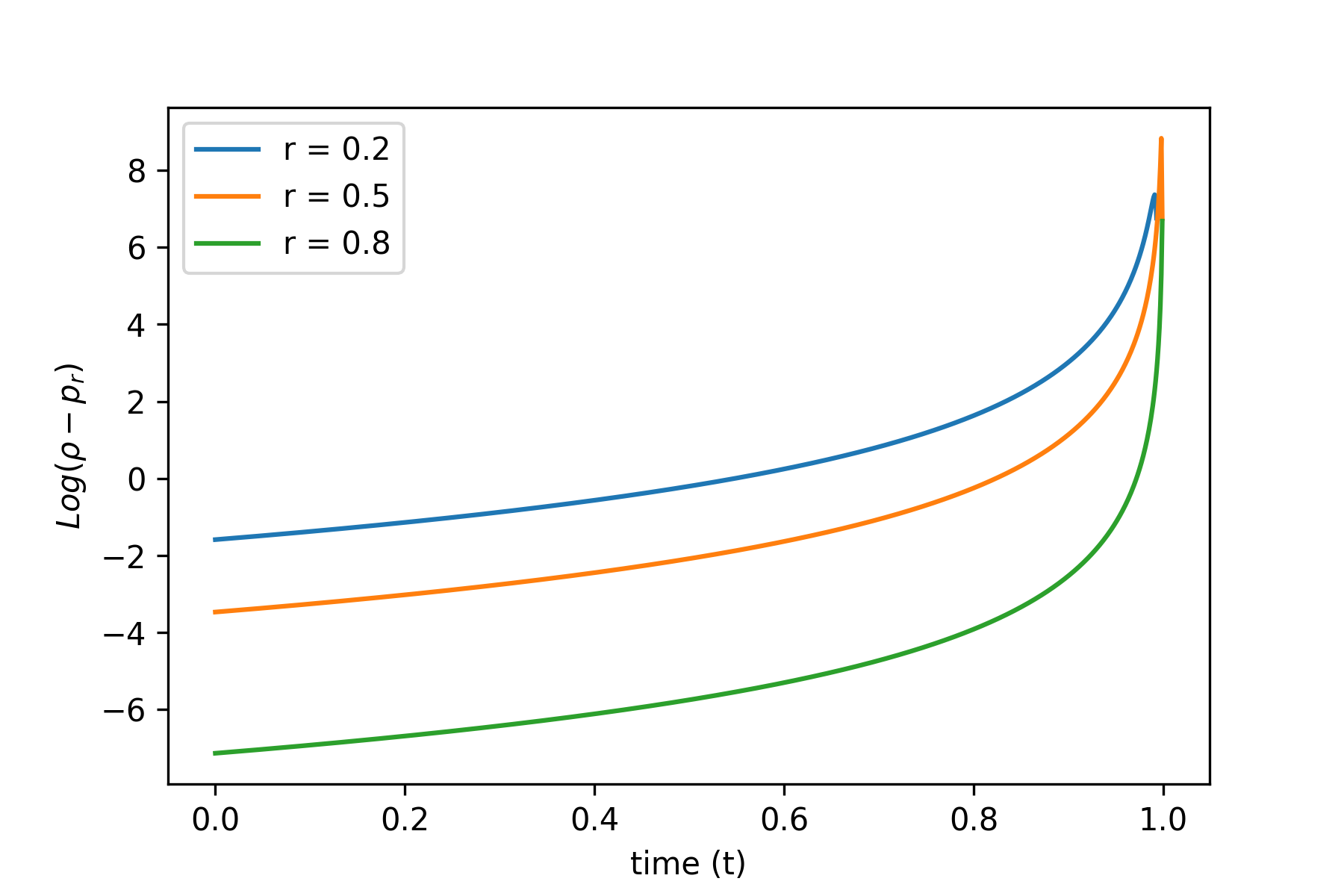}\hfill
\includegraphics[width=.33\textwidth]{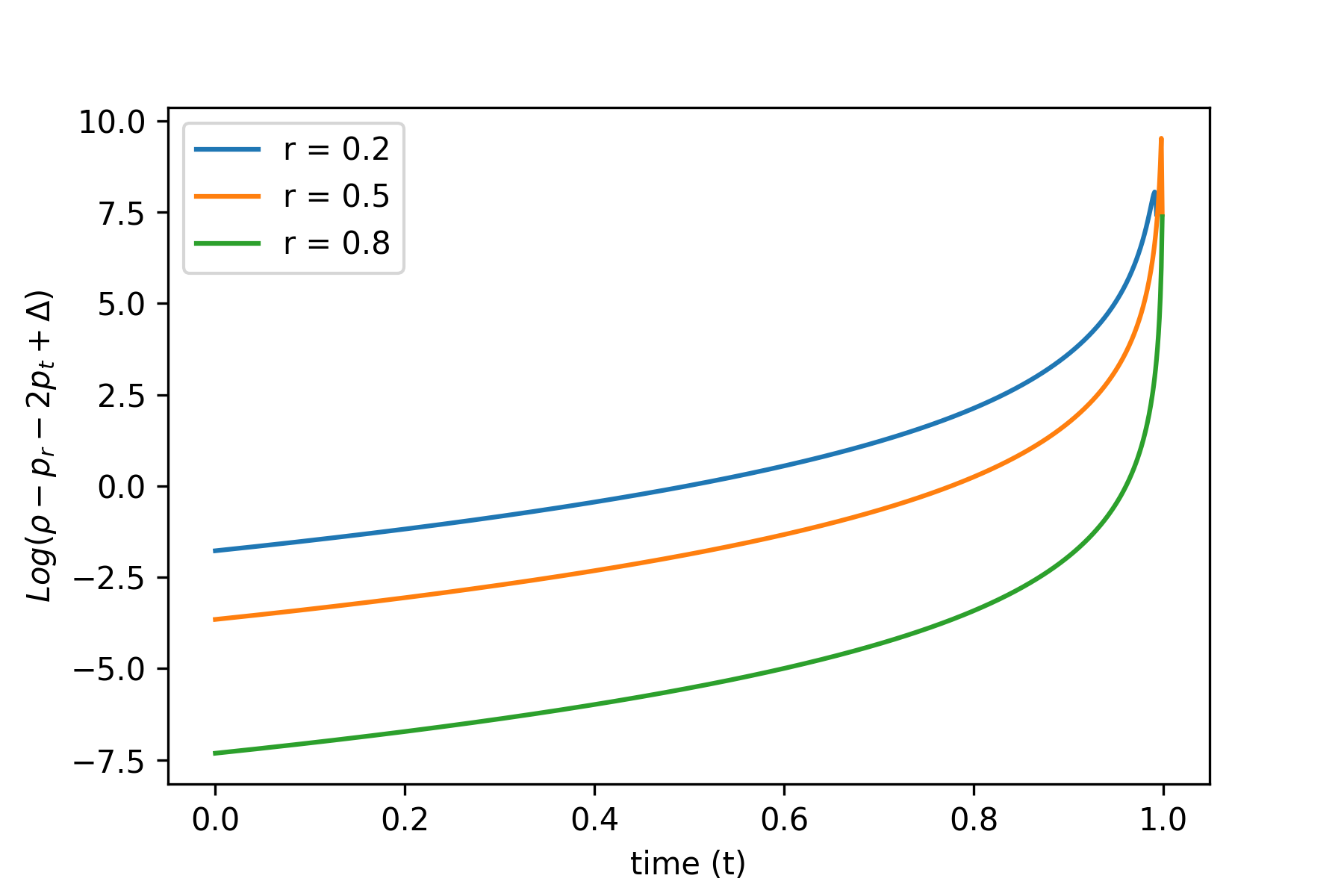}\hfill
\includegraphics[width=.33\textwidth]{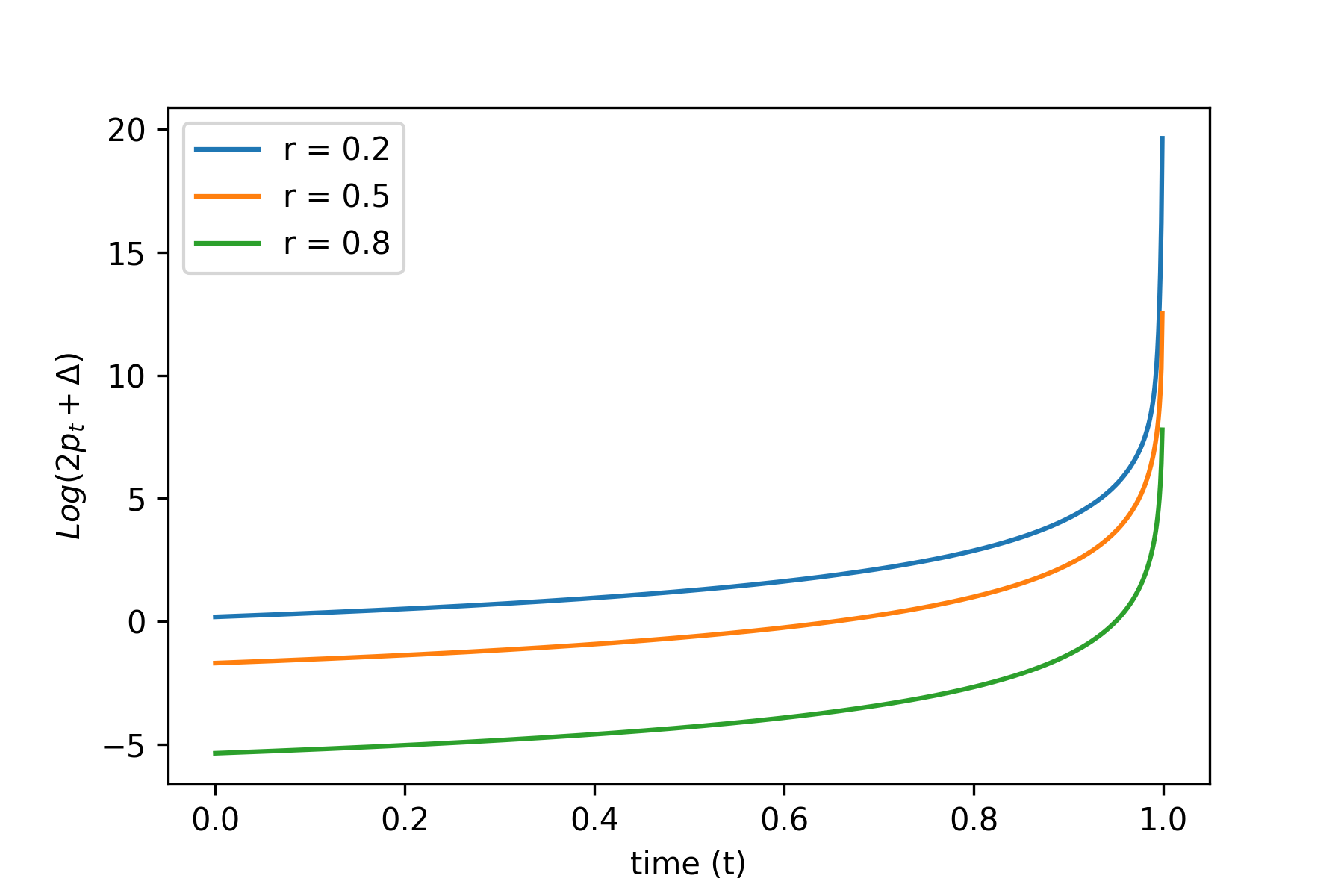}\hfill
\caption{Energy conditions for the model of gravitational collapse with $ f(R) = kR^4 $}
\label{fig3}
\end{figure}

Equations (\ref{53}) to (\ref{56}) represents analytical solutions of gravitational collapse in $ R^4 $ corrected $ f(R) $ gravity. The gravitational collapse starts at $ t = 0 $ and due to choice of $ b = 1 $, this model also governs the end of collapse in singularity at $ t = 1 $. However, due to higher-order correction in the gravitational model, the stronger gravitational pull crunches matter with more power. This can be seen by comparing scales of physical quantities in $ R^2 $ corrected and $ R^4 $ corrected gravity. Equations (\ref{54}) and (\ref{55}) show that anisotropy does not vanish for this case also, as was the case for $ R^2 $ corrected gravity. In the following section, we show that anisotropy has to be taken into account for describing the gravitational collapse of stars in the general $f(R)$ gravity model. We also plot all the energy conditions for this model in FIG \ref{fig3}. The graphs show that this model satisfies all the energy conditions, and thus it is physically acceptable. 

\section{Physical properties of gravitationally collapsing stars}\label{sec7}
In section \ref{sec4}, we have computed the energy density, pressures, and heat flux density of gravitationally collapsing stars. The gravitational collapse in $ f(R) $ gravity also exhibits other useful properties, which are useful to study the inherent nature of such gravitationally collapsing compact stars. Thus, we will study the other useful properties of gravitationally collapsing stars such as shear and anisotropic behavior of matter undergoing gravitational collapse and formation of the apparent horizon by calculating various quantities like 4-velocity, 4-acceleration and expansion parameter of matter, the projection tensor, and shearing tensor. The 4-velocity and the radial vector used in (\ref{3}) for the space-time metric (\ref{5}) and having functions $ y(r) $ and $c(t)$ given by (\ref{33}) are given by

\begin{equation}
\label{57}
    u^{\alpha} = \left( \left( 1 - \frac{r}{a} \right)^2, 0, 0, 0 \right)  \quad\mathrm{and}\quad v^{\alpha} = \left(0, \frac{a\left( 1 - \frac{r}{a} \right)^3}{2\sqrt{2}\sqrt{1-b t}}, 0, 0\right)
\end{equation}

The 4-acceleration of the fluid can be computed from the following equation
\begin{equation}
\label{58}
    a^{\alpha} = u^{\beta}(\nabla_{\beta}u^{\alpha}) 
\end{equation}

which has the following non-zero radial component

\begin{equation}
\label{59}
    a^{1} = \frac{a\left( 1 - \frac{r}{a} \right)^5}{4(1- b t)} 
\end{equation}

Now we computes the expansion parameter as follow
\begin{equation}
\label{60}
   \Theta =  \nabla_{\alpha}u^{\alpha} = -\frac{3b\left( 1 - \frac{r}{a} \right)^2}{2(1-b t)}
\end{equation}

The shear tensor can be computed from the following identity
\begin{equation}
\label{61}
    \sigma_{\alpha \beta} = u_{(\alpha;\beta)} + a_{(\alpha}u_{\beta)} - \frac{1}{3}\Theta(g_{\alpha\beta} + u_{\alpha}u_{\beta})
\end{equation}

where $ h_{\alpha \beta} = g_{\alpha \beta} + u_{\alpha} u_{\beta} $ is known as projection tensor. From equation (\ref{61}), we compute the components of shearing tensor as follow

\begin{equation}
\label{62}
    \sigma_{00} = \frac{-b\left( 1 - \frac{r}{a} \right)^2}{2(1-b t)}\left(-\frac{1}{\left( 1 - \frac{r}{a} \right)^4} + \frac{1}{\left( 1 - \frac{r}{a} \right)^4}\right) = 0
\end{equation}

\begin{equation}
\label{63}
    \sigma_{01} = \sigma_{10} = \frac{1}{a\left( 1 - \frac{r}{a} \right)^3} - \frac{1}{a\left( 1 - \frac{r}{a} \right)^3} = 0
\end{equation}

\begin{equation}
\label{64}
    \sigma_{11} = \frac{2(-\frac{b}{2})(\frac{4}{a^2})}{\left( 1 - \frac{r}{a} \right)^4} - \frac{\left( 1 - \frac{r}{a} \right)^2(-\frac{b}{2})}{(1-b t)}\left( \frac{2(1-b t)\frac{4}{a^2}}{\left( 1 - \frac{r}{a} \right)^6} \right) = 0
\end{equation}

\begin{equation}
\label{65}
    \sigma_{22} = \frac{(-\frac{b}{2})}{\left( 1 - \frac{r}{a} \right)^2} - \frac{(-\frac{b}{2})\left( 1 - \frac{r}{a} \right)^2}{1-b t}\left( \frac{1-b t}{\left( 1 - \frac{r}{a} \right)^4} \right) = 0
\end{equation}

\begin{equation}
\label{66}
    \sigma_{33} = \frac{(-\frac{b}{2})}{\left( 1 - \frac{r}{a} \right)^2} sin^2\theta -\frac{(-\frac{b}{2})\left( 1 - \frac{r}{a} \right)^2}{1-b t}\left( \frac{1-b t}{\left( 1 - \frac{r}{a} \right)^4} \right)sin^2\theta = 0
\end{equation}

Other components of the shearing tensor (\ref{61}) also vanish due to the vanishing of individual parts. Thus, all components including (\ref{62}) to (\ref{66}) vanishes. Thus, the prior assumption of vanishing shear components holds for the study. It is important to note that the vanishing shear is only possible if gravitationally collapsing stars are being studied in the separable form of the interior space-time metric. The gravitational collapse of compact stars described by the non-separable form of interior metric involves the shear, and thus it has a non-zero shearing tensor. The detailed study of the gravitational collapse in the non-separable form of interior metric considering shear and its evolution is given in the reference \cite{2008GReGr..40.2149P}.
\par Now, we compute the anisotropy $ (S) $ of the stars undergoing the gravitational collapse described by the modified $ R + f(R) = R + k R^m$ gravity model under consideration. From equation (\ref{7}) and (\ref{8}) we have the following identity,

\begin{equation}
\label{67}
    S = p_r - p_t = \frac{f_R'(y'^2+y y'' )}{2y y'^3 c^2} - \frac{f_R''}{2c^2y'^2}
\end{equation}

For the model $ f(R) = k R^m $, the values of $ f_R' $ and $ f_R'' $ is given by

\begin{equation}
\label{68}
    f_R' = -k(2m)(m-1)R^{m-1}\frac{y'}{y}
\end{equation}

\begin{equation}
\label{69}
    f_R'' = k R^{m-1}\left[2m(m-1)(2m-1)\frac{y'^2}{y^2} - 2m(m-1) \frac{y''}{y}\right]
\end{equation}

Using equations (\ref{68}) and (\ref{69}) in equation (\ref{67}), we compute the anisotropy as follow

\begin{equation}
\label{70}
    S = p_r - p_t = k(2m^2)(m-1)R^m 
\end{equation}

The anisotropy $ S $ can be written in terms of space-time coordinate by utilizing equation (\ref{34}) in (\ref{70}) as follow

\begin{equation}
\label{71}
    S = p_r - p_t = k(-1)^m(2m^2)(m-1)\frac{\left( 1 - \frac{r}{a} \right)^4m}{(1-b t)^m} 
\end{equation}

Equation (\ref{70}) shows that the anisotropy can vanish for $ k = 0 $ only, in which case the solutions will reduce to GR. This fact can also be observed by equations (\ref{45}) to (\ref{48}). Thus, for the case of modified $ f(R) $ gravity, we have to consider the anisotropic behavior of gravitationally collapsing stars described by the interior metric (\ref{5}). Equation (\ref{71}) shows that anisotropy does not vanish at the center of the gravitationally collapsing star, instead of vanishing central anisotropy for the stable, compact stars. Thus, it shows that extra curvature source (\ref{4}) in $ f(R) $ gravity forces the matter distribution to be anisotropic in nature for gravitationally collapsing stars described by the interior space-time (\ref{5}). 
\par Now we investigate the formation of the apparent horizon and singularity formation for gravitational collapse of compact stars with the model of $ f(R) $ gravity under consideration. From equations (\ref{33}) and (\ref{34}), it can be seen that the gravitational collapse ends in singularity at $ t = \frac{1}{b} $ when $ c(t) = 0 $ and Ricci scalar diverges at all co-moving radii. Now the question is that whether the singularity is naked or the horizon is formed before the observers can see the formation of a singularity. In the second case, there will be a black hole when the horizon is formed while gravitational collapse continues to the formation of a singularity. Whether the singularity will be naked or not can be investigated by solving the equation of formation of an apparent horizon as follow

\begin{equation}
    \label{72}
    g^{\alpha \beta}\partial_{\alpha}(c(t)^2y(r)^2)\partial_{\beta}(c(t)^2y(r)^2) = 0
\end{equation}

We use the general forms of $ y(r) $ and $ c(t) $ as obtained in equation (\ref{33}) and solve equation (\ref{72}) as follow

\begin{equation}
    \label{73}
    t_h = \frac{4- 2b^2}{4b}
\end{equation}

Where $ t_h $ is the time at which horizon is formed. For $ b = 1 $, the time when horizon is formed is $ t_h = \frac{1}{2} $. Thus, for the particular solution (\ref{33}) of interior space-time metric describing gravitational collapse with $ b = 1 $, the horizon is formed at $ t = \frac{1}{2} $. The particular solution with $ b = 1 $ is of interest because it describes completely physical collapse satisfying all energy conditions as well as investigated in section \ref{6}    for gravitational collapse model $GR$, $ R+k R^2 $ and $ R+k R^4 $ gravity. In that case, all observers will see the formation of the horizon before the singularity forms, and thus singularity will be hidden behind the horizon. Thus, observers for those particular cases of gravitational collapse will see that gravitational collapse is started at $ t = 0 $ and horizon is formed at $ t = \frac{1}{2} $, before the formation of singularity at $ t = 1 $. However, it should be noted that it is not the only possible physically acceptable solution. We have chosen this class of solutions with $ b = 1 $. However, for different classes of solutions satisfying all energy conditions with different values of $ b $ is maybe possible. For different solutions, the horizon and singularity will be formed at different times or gravitational collapse may also end in naked singularity as can be investigated from equation (\ref{73}).

\section{Discussion}
\label{sec8}
In this paper, we obtained exact solutions describing gravitational collapse of compact stars in $ R + f(R) = R + k R^m $ gravity. We solved field equations of gravitational collapse in $ f(R) $ gravity by considering junction conditions and implementing the method of $ R $ matching to ensure the continuity of the Ricci scalar and its derivatives across the matching hypersurface. These extra junction conditions restrict the physically possible solutions greatly, and thus, there are not many exact solutions are available for the problem in the literature. Recently some exact solutions have been developed by some authors, but the problem is considered only for the restricted class of $ f(R) $ model. 
\par Thus, in this paper, we formulated the exact solutions describing the gravitational collapse of compact stars in general and physically important model of $ f(R) $ gravity. We then investigated the values of model parameters such that they satisfy all necessary energy conditions. We have demonstrated the model's validity for $GR$, $ R+k R^2 $ and $ R+k R^4 $ gravity. However, the model is utterly physical for any order of corrections in $ f(R) = k R^m $ gravity. We have presented a comprehensive graphical analysis of energy conditions for each case and explained all case and their consequences. We also explained how the gravitational collapse scenario is different for $ GR $ and lower and higher-order correction in $ f(R) $ gravity. For example, in the $ GR $ case, the anisotropy of gravitationally collapsing stars vanishes identically for all time at different radii, in the case of the particular interior metric considered here. However, for modified $ f(R) $ gravity, the anisotropy of stars has to be considered.
\par After obtaining a physically acceptable and well-behaved gravitational collapse model, we presented some important physical properties of gravitationally collapsing stars. We derived various quantities like 4-velocity, 4-acceleration, and the expansion parameter of the matter undergoing gravitational collapse. Then we calculated the shearing tensor and anisotropy of compact stars described by the particular interior metric considered in this paper. From that, we concluded that separable-form of interior metric leads to vanishing shear inside the collapsing stars. However, anisotropy has to be considered for the model of $ f(R) $ gravity investigated in this paper. Finally, we studied the formation of singularity and horizon for stars undergoing gravitational collapse. From that, we concluded that the model studied in this paper leads the gravitational collapse of compact stars to end in a black hole by forming a horizon before the formation of singularity at the end of the collapse.

\section*{Acknowledgement}
The author is very grateful to Professor S. Odintsov for his comments and insightful suggestions throughout the
research work.

\bibliographystyle{unsrt}
\bibliography{main}

\end{document}